\documentclass[11pt]{article}
\usepackage{amsmath}
\usepackage{amssymb}
\usepackage{graphicx}
\usepackage{enumerate}
\usepackage{cite}
\usepackage{color}
\usepackage[margin=1.0in]{geometry}
\usepackage{authblk}
\normalsize \baselineskip 16pt
\newcommand {\beq} {\begin{equation}}
\newcommand {\eeq} {\end{equation}}

\begin{document}

\title{When discrete fronts and pulses form a single family: FPU chain with hardening-softening springs}
\author[1]{Anna Vainchtein}
\author[2]{Lev Truskinovsky}
\affil[1]{\small Department of Mathematics, University of Pittsburgh, Pittsburgh, Pennsylvania 15260, USA, \texttt{aav4@pitt.edu}}
\affil[2]{\small PMMH, CNRS--UMR 7636, ESPCI ParisTech, 10 Rue Vauquelin, Paris, 75005, France \texttt{lev.truskinovsky@espci.fr}}

\maketitle

\begin{abstract}

We consider a version of the classical Hamiltonian FPU (Fermi-Pasta-Ulam) problem with nonlinear force-strain relation in which a hardening response is taken over by a softening regime above a critical strain value. We show that in addition to pulses (solitary waves) this discrete system also supports non-topological and dissipation-free fronts (kinks). Moreover, we demonstrate that these two types of supersonic traveling wave solutions belong to the same family. Within this family, solitary waves exist for continuous ranges of velocity that extend up to a limiting speed corresponding to kinks. As the kink velocity limit is approached from above or below, the solitary waves become progressively more broad and acquire the structure of a kink-antikink bundle. Direct numerical simulations and Floquet analysis of linear stability suggest that all of the obtained solutions are effectively stable. To motivate and support our study of the discrete problem we also analyze a quasicontinuum approximation with temporal dispersion. We show that this model captures the main effects observed in the discrete problem both qualitatively and quantitatively.

\end{abstract}

\section{Introduction}

Front-shaped kinks and pulse-shaped solitary waves are usually perceived  as two fundamentally different types of traveling waves that are ubiquitous in nonlinear discrete systems. Both kinks and solitary waves are localized coherent structures that represent far-from equilibrium collective phenomena emerging from the underlying many-body interactions. They are encountered in integrable and non-integrable Hamiltonian systems and can be stable or unstable. Together with breathers, they play an important role as building blocks in complex dynamic patterns in nonlinear systems and contribute crucially to the mechanical energy transmission at the microscale \cite{speight1999topological,malomed2020nonlinearity,askari2020collision,kivshar1989dynamics}.
Important applications associated with mechanical kinks and solitary waves are mitigation  of  impact loadings,  transmission, guiding and encryption of mechanical information, including enabling logic operations and activating soft robotics
\cite{bertoldi2017flexible,yasuda2019origami}.

Kinks, originally introduced  in the context of  sine-Gordon-type equations,  are usually perceived as self-induced topological defects   representing connections between  different  energy wells of a potential. The dynamics of discrete kinks is typically  dissipative  due to  radiative losses  which results in a finite driving force needed for such defects  to be spatially displaced \cite{braun2004frenkel,peyrard1984kink,malomed1990domain}. Kinks and antikinks correspond to the discrete spectrum of a nonlinear eigenvalue problem defining their velocity. They are described by heteroclinic trajectories of the corresponding differential equations and move with specific, usually subsonic speeds \cite{kawasaki1982kink,pismen2006patterns, pego1989front, rubinstein2004detachment}. Typical mechanical examples of kinks (not to be confused with shock waves in Burgers-type equations) are phase boundaries \cite{truskinovsky1993kinks} and dislocations \cite{fitzgerald2016kink}. In contrast, solitary waves, often discussed in the context of KdV-type equations, can be characterized as localized, non-topological and usually non-dissipative wave packets whose existence does not require a multiwell structure of the potential \cite{remoissenet2013waves, newell1985solitons, fokas2012important,Vainchtein22}. Solitary waves usually  move with supersonic speeds and belong to a continuous spectrum of traveling wave solutions described by homoclinic trajectories in the phase space \cite{ablowitz2011nonlinear}. Typical mechanical examples are tidal bores \cite{apel2002oceanic} and self-healing pulses imitating earthquakes \cite{pomyalov2023self}.

While both kinks and solitary waves first appeared in the context of integrable, exactly solvable nonlinear models, which describe   physical systems only within a certain approximation, here we consider a more realistic non-integrable Hamiltonian  mechanical system that bears both kinks and solitary waves. More specifically, we consider the well known discrete Fermi-Pasta-Ulam (FPU) model
\cite{fermi1955studies,gallavotti2007fermi,berman2005fermi}
and in this way address the issue of the coexistence of kinks and solitary waves  in a one-dimensional mass-spring chain. Such coexistence was absent in the $\alpha$-FPU setting, which relied on quadratic nonlinearity of the force-strain relation. Here we consider an extension  of this  prototypical model in which a hardening response is taken over by a softening regime above a critical strain value. Meanwhile, the interaction potential is a \emph{convex} function of strain in the relevant strain interval.

The choice of \emph{hardening-softening}  interactions is inspired by stress-strain laws in a range of soft biological tissues  from  skin  to muscles \cite{Yasenchuk21}.  For instance, in tendons and  ligaments the hardening stage of the  mechanical  response can  be linked to the  straightening of crimped collagen fibers while the softening  stage may be due to the beginning of the distributed microscopic fracturing of these fibers \cite{Yasenchuk21,Sensini18}.
Hardening to softening transition is also ubiquitous in elastomeric molecular composites \cite{Millereau18} and can be even mimicked in NiTi mesh implants \cite{Yasenchuk21}.  Note that apparently similar convex material response but of
\emph{softening-hardening} type have been studied before in continuum setting, however, with the dynamic response found to be uneventful and basically the same as in the other well studied cases with double-well potentials, where kinks that are topological are fundamentally different from solitary waves that are non-topological \cite{knowles2002impact, schulze1997undercompressive}.

Existence and properties of either kinks or solitary waves in hardening-softening discrete FPU system have been studied before  \cite{Iooss00,HerrmannRademacher10,Herrmann11,Gorbushin20,Gorbushin21} but the unifying perspective on their coexistence and interplay  in a generic setting was missing.
Other systems supporting coexisting kinks and solitary waves include discrete transmission lines
\cite{kogan2022kinks,kogan2024kinks},
complex  Ginzburg-Landau equation \cite{hakim1990fronts, van1990pulses, malomed1990kinks} and the Gardner equation
\cite{RosenauOron20,RosenauPikovsky20,RosenauPikovsky21}.
Studies specifically focused on  the  interrelation between kinks and solitary waves in these and other related systems include \cite{duan1995fronts,kazantsev1997pulses,chang1996local,triki2022dark, fochesato2005generalized,YeeChow10,chang2008multiplicity,kim2000soliton,zheng2023solitary}.

In the present  paper we clarify why in addition to conventional solitary waves,  the hardening-softening  discrete FPU model also necessarily supports non-topological and dissipation-free kinks. Kinks have polarity and thus always  arrive together with their twins of opposite polarity, which we call antikinks. Under certain conditions  kinks and antikinks can form a bundle and in this way annihilate  their polarity.   Since  kink and antikink can move with the same speed, the bundled compact configurations can also move with a constant speed. The ensuing ``marginal'' solitary waves lie on the boundary of the solitary-wave domain in the space of parameters (the kink limit).

More specifically, we show that kinks and solitary waves, viewed as two types of discrete supersonic traveling wave solutions of the FPU model, belong to the same family. Within this family, solitary waves exist for continuous ranges of velocity that extend up to a limiting speed corresponding to kinks. As the kink velocity limit is approached from above or below, the solitary waves become progressively more broad and acquire the structure of a kink-antikink bundle. This scenario differs from the systems with nonconvex interaction potentials, where in contrast to non-topological and non-dissipative solitary waves, the generic kinks are necessarily topological and dissipative (radiative).

In the recent paper \cite{Gorbushin22} we showed that in the Hamiltonian FPU model there can appear exactly three distinct classes of steady switching fronts, subkinks, shocks and superkinks, which fundamentally differ in how (and whether) they produce and transport oscillations. In this classification subkinks are subsonic and dissipative (radiative), shocks are supersonic and dissipative and superkinks are supersonic and non-dissipative. The kinks considered in the present paper are supersonic and non-dissipative and thus are \emph{superkinks} in the sense of \cite{Gorbushin22}.

After formulating the discrete problem and providing the conditions necessary for the coexistence of superkinks and solitary waves, we consider a quasicontinuum (QC) approximation of the discrete problem that adds to the conventional continuum elastrodynamics a mixed space-time higher-order derivative term describing temporal dispersion and accounting for microinertia contribution to the kinetic energy \cite{collins1981quasicontinuum,Rosenau86,kevrekidis2002continuum,feng2004quasi}.
In contrast to the more conventional QC models that involve purely spatial dispersion term \cite{christov2007boussinesq,  kunin2012elastic,charlotte2008towards}, this approximation generates a bounded dispersion relation for a linearized problem, which precludes short-wave instabilities. We present a detailed analysis of the QC problem and show that it possesses a family of superkink and solitary wave solutions, which are computed explicitly for a cubic extension of the $\alpha$-FPU interaction force. The analytical transparency of the QC model allows one to understand in full detail the singular role played by superkinks embedded inside the continuous range of solitary waves and to associate the special kinetic relation, characterizing such kinks, with their non-dissipative nature.

Using the obtained solutions of the QC problem as a starting point, we then proceed to compute the corresponding traveling wave solutions of the discrete problem. To do so, we take advantage of the fact that traveling waves are periodic modulo shift by one lattice space and thus can be computed as fixed points of the corresponding nonlinear map \cite{Aubry09,Vainchtein20,James21}. We then follow the approach in \cite{Cuevas17,Xu18,Vainchtein20} and exploit the periodicity-modulo-shift of the traveling waves to study their linear stability by computing the Floquet multipliers associated with the corresponding linearized problem. Similar to other related problems \cite{Marin98,Xu18}, our Floquet analysis indicates mild oscillatory instabilities that appear to be a spurious artifact of the chain size in the computations, since their magnitude decreases for longer chains. Effective stability of the computed waves is supported by direct numerical simulations that show their steady propagation. We also present some simulation results for initial value problems that show formation and steady motion of superkinks and solitary waves.

Comparison of the computed solutions of the discrete problem with the corresponding exact solutions of the QC model shows a very good agreement on both qualitative and quantitative levels. It is important to mention that the proposed QC framework not only provides a transparent interpretation of the two types of nonlinear waves, but also helps to explain in physical terms why kinks are
dissipation-free and why at least some solitary waves can be viewed as nonlinear superpositions of kinks and antikinks. Previous results for this problem, revealing similar effects, concern a  bilinear version  of the  model which turns out to be analytically solvable in both discrete and QC versions \cite{Gorbushin20, Gorbushin21}.

The paper is organized as follows. In Section~\ref{sec:TWgen} we introduce the discrete problem and discuss some general properties of superkinks and solitary waves. In Section~\ref{sec:QC} we introduce the QC model and present the phase-plane analysis of the problem for general hardening-softening nonlinearity. Explicit traveling wave solutions of the QC problem with a cubic nonlinearity are derived and discussed in Section~\ref{sec:QCsolns}. These solutions are used in Section~\ref{sec:Dsolns} to obtain the traveling wave solutions of the discrete problem as fixed points of the corresponding nonlinear map and compare the results of the two problems. Stability of the obtained solutions of the discrete problem is investigated in Section~\ref{sec:stab} using both Floquet analysis and direct numerical simulations. Concluding remarks are presented in the final Section~\ref{sec:conclusions}.

\section{Supersonic kinks and solitary waves: general properties}
\label{sec:TWgen}
\begin{figure}
\centering
\includegraphics[width=0.5\textwidth]{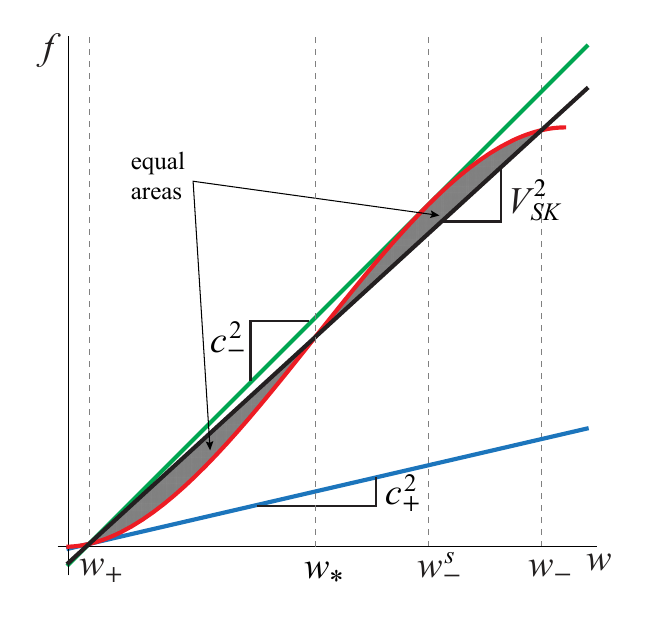}
\caption{\footnotesize Interaction force $f(w)$ (red) and the Rayleigh line (black) connecting $(w_{+},f(w_{+}))$ and $(w_{-},f(w_{-}))$ (black). The strain $w_*$ marks the transition from the hardening (convex) to softening (concave) regime. A superkink transition wave with limiting states $w_{\pm}$ and supersonic velocity $V_{SK}$ such that $V_{SK}^2>f'(w_{\pm})$ exists when the two shaded areas cut by the Rayleigh line, which has the slope $V^2$, are equal. The associated solitary wave solutions have velocity $V$ such that either $c_{+}^2<V^2<V_{SK}^2$ or $V_{SK}^2<V^2<c_{-}^2$, where $c_+^2$ and $c_{-}^2$ are the slopes of the blue and green straight lines passing through $(w_{+},f(w_{+}))$ and tangent to $f(w)$ at $w_+$ and $w_{-}^s$, respectively. See the text for more details.}
\label{fig:superkink}
\end{figure}

We consider the basic FPU model, which describes the dynamics of a one-dimensional chain of identical masses interacting with their nearest neighbors. The dimensionless governing equations are
\beq
\ddot{u}_n=f(u_{n+1}-u_n)-f(u_n-u_{n-1}),
\label{eq:FPU}
\eeq
where $u_n(t)$ is the displacement of $n$th particle at time $t$, $\ddot{u}_n(t)=u_n''(t)$, and $f(w)=\Phi'(w)$ is the nonlinear interaction force obtained from the interaction potential $\Phi(w)$. Introducing particle velocities $v_n=\dot{u}_n(t)=u_n'(t)$ and strain variables $w_n=u_n-u_{n-1}$, we can rewrite \eqref{eq:FPU}
as the first-order system
\beq
\dot{w}_n=v_n-v_{n-1}, \quad \dot{v}_n=f(w_{n+1})-f(w_n).
\label{eq:FPUsystem}
\eeq
Written in terms of strain variables alone, the equations are
\beq
\ddot{w}_n=f(w_{n+1})-2f(w_n)+f(w_{n-1}).
\label{eq:FPUstrain}
\eeq
In what follows, we assume that $f(0)=0$ and that in an interval $(\alpha,\beta)$ of strains that includes zero, we have $f'(w)>0$. Note that this implies that the corresponding interaction potential $\Phi(w)=\int_0^w f(s)ds$ is convex in the interval
$(\alpha,\beta)$. We further assume that there exists $w_*$ such that $0<w_*<\beta$, $f''(w_*)=0$, and we have $f''(w)>0$ for $\alpha<w<w_*$ (\emph{hardening response}) and $f''(w)<0$ for $w_*<w<\beta$ (\emph{softening response}), so that $f'(w)$ has a local maximum at $w=w_*$. A simple example of such hardening-softening (convex-concave) interaction force is the cubic interaction force
\beq
f(w)=aw^3+bw^2+w, \quad a<0, \quad b>0,
\label{eq:cubic}
\eeq
shown by the red curve in Fig.~\ref{fig:superkink}. In this case we have
\beq
(\alpha,\beta)=\left(-\dfrac{\sqrt{b^2+3|a|}-b}{3|a|},\;\dfrac{b+\sqrt{b^2+3|a|}}{3|a|}\right), \quad w_*=\dfrac{b}{3|a|}.
\label{eq:cubic_more}
\eeq
We reiterate that while the convex, hardening part of $f(w)$ can be associated with reorganization of the micro-constituents contributing to interconnectivity and increasing rigidity, the concave, softening part can be linked to the loss of interconnectivity associated with the ultimate emergence of damage \cite{Yasenchuk21,Sensini18,Millereau18}. In what follows, we restrict our attention to solutions with strain values in the $(\alpha,\beta)$ interval given in \eqref{eq:cubic_more}, where the corresponding potential $\Phi(w)$ is convex, thus preventing the non-physical behavior at large $|w|$.

We are interested in traveling waves that connect stable equilibrium states of the system, with constant strains $w_\pm$ such that $f'(w_{\pm})>0$ and constant particle velocities $v_{\pm}$, and propagate with velocity $V$ that is supersonic with respect to both limiting states: $V^2>f'(w_{\pm})$. Thus, we seek solutions in the form
\beq
w_n(t)=w(\xi), \quad v_n(t)=v(\xi), \quad \xi=n-Vt,
\label{eq:TWansatz}
\eeq
where
\beq
\lim_{\xi \to \pm \infty} w(\xi) = w_{\pm}, \quad \lim_{\xi \to \pm \infty} v(\xi) = v_{\pm}.
\label{eq:BCs}
\eeq
Monotone traveling fronts connecting two \emph{different} states, $w_{+} \neq w_{-}$, correspond to \emph{superkinks}. As shown in \cite{Iooss00}, in the case of smooth $f(w)$ small-amplitude superkinks bifurcate from local maxima of $f'(w)$ connecting convex and concave parts of $f(w)$. Global existence of such fronts in the FPU problem with convex-concave nonlinearity was established in \cite{HerrmannRademacher10,Herrmann11} under the area condition discussed below. As we will show, superkinks are closely related to \emph{solitary waves}, pulse-like solutions of \eqref{eq:TW_discrete} connecting \emph{identical} limiting states, $w_{-}=w_{+}=w_B$, and propagating with supersonic velocities. Existence of such solutions has been shown in \cite{Friesecke94}. In what follows, we focus on these two types of traveling waves.

For both types of solutions, the function $w(\xi)$ must satisfy the advance-delay differential equation
\beq
V^2w''(\xi)=f(w(\xi+1))-2f(w(\xi))+f(w(\xi-1))
\label{eq:TW_discrete}
\eeq
obtained by substituting \eqref{eq:TWansatz} into \eqref{eq:FPUstrain}. Using \eqref{eq:FPUsystem} instead, we obtain the equivalent system of first-order equations:
\beq
-Vw'(\xi)=v(\xi)-v(\xi-1), \quad -Vv'(\xi)=f(w(\xi+1))-f(w(\xi)).
\label{eq:TWsystem_discrete}
\eeq
Combining these two equations, we obtain the energy balance law \cite{HerrmannRademacher10}
\beq
-V\dfrac{d}{d\xi}\left[\dfrac{1}{2}v^2(\xi)+\Phi(w(\xi))\right]=f(w(\xi+1))v(\xi)-f(w(\xi))v(\xi-1).
\label{eq:TWenergy_discrete}
\eeq
Integrating the equations in \eqref{eq:TWsystem_discrete} over the finite interval $[-N,N]$ and taking the limit $N \to \infty$ as in \cite{HerrmannRademacher10} (see also \cite{Serre07,Aubry09}), we recover the classical Rankine-Hugoniot jump conditions
\beq
-V(w_{+}-w_{-})=v_{+}-v_{-}, \quad -V(v_{+}-v_{-})=f(w_{+})-f(w_{-}),
\label{eq:RHconds}
\eeq
which upon the elimination of $v_{\pm}$ yield the single condition
\beq
f(w_{+})-f(w_{-})=V^2(w_{+}-w_{-}).
\label{eq:Rline}
\eeq
This condition trivially holds for solitary waves, since $w_{-}=w_{+}$ in that case. For superkinks, it states that the slope of the \emph{Rayleigh line} connecting $(w_+,f(w_+))$ and $(w_{-},f(w_{-}))$ equals $V^2$, as shown in Fig.~\ref{fig:superkink}.

Similarly, integrating \eqref{eq:TWenergy_discrete}, we obtain
\beq
-V\left(\dfrac{1}{2}v_+^2+\Phi(w_+)-\dfrac{1}{2}v_{-}^2-\Phi(w_{-})\right)=f(w_+)v_+-f(w_{-})v_{-}.
\label{eq:energy_jump}
\eeq
A simple calculation then shows that \eqref{eq:energy_jump}, \eqref{eq:Rline} and the first of \eqref{eq:RHconds} imply \cite{HerrmannRademacher10,Herrmann11}
\beq
\Phi(w_{+})-\Phi(w_{-})-\dfrac{1}{2}(w_{+}-w_{-})(f(w_{+})+f(w_{-}))=0.
\label{eq:zero_G}
\eeq
For solitary waves, this condition is again trivially satisfied. For superkinks, however, the condition \eqref{eq:zero_G} has an important physical meaning. It is a \emph{kinetic relation} that states that the \emph{driving force} $G=\Phi(w_{+})-\Phi(w_{-})-\frac{1}{2}(w_{+}-w_{-})(f(w_{+})+f(w_{-}))$ \cite{Trusk87} on the moving front is zero, and thus
there is no dissipation associated with its motion. Geometrically, this means that the two areas cut by the Rayleigh line from $f(w)$ must be equal, as shown in Fig.~\ref{fig:superkink}.

Conditions \eqref{eq:Rline} and \eqref{eq:zero_G} are thus necessary for the existence of a superkink solution. Therefore, they appear  in the existence conditions obtained  \cite{HerrmannRademacher10,Herrmann11} and  were also independently obtained in the case of piecewise linear $f(w)$  in  \cite{Gorbushin20,Gorbushin21,Gorbushin22},  where they were  linked to the absence of elastic radiation of lattice waves which serve as a Hamiltonian analog of macroscopic dissipation. Importantly, the two conditions imply that in the case of superkinks, only one of the values $w_{-}$, $w_+$ and $V$ can be prescribed independently. In particular, they determine $w_{\pm}$ as a function of $V$.

Note that for each superkink solution propagating with velocity $V$, there exists a solution of the same form but velocity $-V$.
In addition, for each kink solution with $w_{-}>w_{+}$, i.e., a front with $w'(\xi)<0$, there is an \emph{antikink} solution with the limiting states interchanged, so that $w'(\xi)>0$, and the same velocity. Meanwhile, solitary waves can be \emph{tensile}, $w(\xi)>w_B$, or \emph{compressive}, $w(\xi)<w_B$. Similar to the superkinks, for each solitary wave moving with velocity $V$, there is a wave of the same form moving with velocity $-V$. Properties of solitary waves associated with a superkink are discussed in detail below.

\section{Quasicontinuum model}
\label{sec:QC}
To motivate and support our study of the discrete problem, we first consider its quasicontinuum (QC) approximation. To obtain it, we note that in Fourier space \eqref{eq:TW_discrete} becomes
\[
k^2 V^2 W(k)=4\sin^2(k/2)F(k),
\]
where $k$ is the wave number, and $W(k)$ and $F(k)$ are the Fourier transforms of $w(\xi)$ and $f(w(\xi))$, respectively. Using the $(2,2)$ Pad\'e approximation, $4\sin^2(k/2) \approx k^2/(1+k^2/12)$, of the discrete Laplacian in Fourier space and taking the inverse Fourier transform, we obtain
\beq
V^2 w''-\dfrac{V^2}{12}w''''=(f(w))''.
\label{eq:TW_QC_gen1}
\eeq
The same traveling wave equation can be obtained by differentiating the regularized Boussinesq partial differential equation
\[
u_{tt}-\dfrac{1}{12}u_{xxtt}=(f(u_x))_x,
\]
which describes the QC model derived in \cite{Rosenau86}, with respect to $x$ and seeking solutions in the form $y(x,t)=u_x(x,t)=w(\xi)$, $\xi=x-Vt$. The above equation can also be derived from the Lagrangian density
\beq
\mathcal{L}=\dfrac{1}{2}\left(u_t^2+\dfrac{1}{12}u_{tx}^2\right)-\Phi(u_x),
\label{eq:Lagrangian}
\eeq
which contains a ``microkinetic'' energy term $(1/24)u_{tx}^2$ in addition to the classical kinetic and potential energy terms.
Integrating \eqref{eq:TW_QC_gen1} twice and using the boundary condition for $w(\xi)$ at $\xi \to \infty$ in \eqref{eq:BCs}, we obtain
\beq
-\dfrac{V^2}{12}w''+V^2 w-f(w)=V^2 w_{+}-f(w_+).
\label{eq:TW_QC_gen2}
\eeq
Applying the boundary condition for $w(\xi)$ at $\xi \to -\infty$ in \eqref{eq:BCs} to \eqref{eq:TW_QC_gen2}, we recover the  Rankine-Hugoniot condition \eqref{eq:Rline}. Integrating \eqref{eq:TW_QC_gen2} and taking into account the boundary condition for $w(\xi)$ at $\xi \to \infty$ in \eqref{eq:BCs} yields
\beq
-\dfrac{V^2}{24}(w')^2=\Phi(w)-\Phi(w_{+})-f(w_{+})(w-w_{+})-\dfrac{V^2}{2}(w-w_{+})^2.
\label{eq:TW_QC_gen3}
\eeq
Applying the boundary condition for $w(\xi)$ at $\xi \to -\infty$ in \eqref{eq:BCs} to \eqref{eq:TW_QC_gen3}, we obtain
\[
\Phi(w_{-})-\Phi(w_{+})-f(w_{+})(w_{-}-w_{+})-\dfrac{V^2}{2}(w_{-}-w_{+})^2=0,
\]
which together with \eqref{eq:Rline} implies \eqref{eq:zero_G}.

To construct a superkink solution of \eqref{eq:TW_QC_gen2}, it thus suffices to find $w_{\pm}$ satisfying \eqref{eq:Rline} and \eqref{eq:zero_G} for a given $V$ (or, equivalently, $w_{-}$ and $V$ satisfying these conditions for a given $w_{+}$) and then solve the first-order equation \eqref{eq:TW_QC_gen3}.

\begin{figure}[t]
\centering
\includegraphics[width=\textwidth]{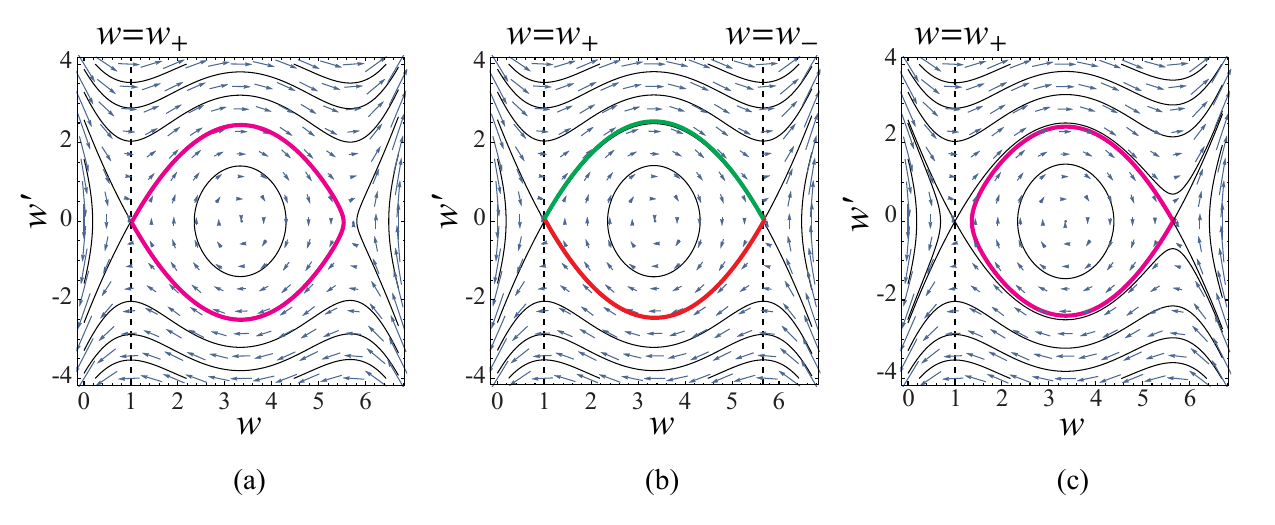}
\caption{\footnotesize Phase portraits for \eqref{eq:TW_QC_gen2} with $f(w)$ given by \eqref{eq:cubic}, $w_+=1$ and (a) $V=V_{SK}-0.001$; (b) $V=V_{SK}$; (c) $V=V_{SK}+0.005$, where $V_{SK}$ satisfies the conditions \eqref{eq:Rline}, \eqref{eq:zero_G} with the corresponding $w_{-}$. See the text for details. Here $V_{SK}=2\sqrt{65}/3$, $w_{-}=17/3$. The colored trajectories
corresponds to a tensile solitary waves in (a), superkinks in (b) and a compressive solitary wave in (c).}
\label{fig:pplane}
\end{figure}
Emergence of superkinks and the associated solitary wave solutions can already be seen from the phase plane analysis of \eqref{eq:TW_QC_gen2} for fixed $w_+<w_*$ and different values of $V$, as illustrated in Fig.~\ref{fig:pplane} for the cubic case \eqref{eq:cubic}. Let $V_{SK}$ denote the velocity of the superkink, which satisfies the conditions \eqref{eq:Rline}, \eqref{eq:zero_G} with the corresponding $w_{-}$. Observe that the Rayleigh line $R(w)=f(w_+)+V^2(w-w_+)$ passing through $(w_+,f(w_+))$ is tangent to $f(w)$ at $w=w_+$ when $V^2=c_+^2$, where
\beq
c_+=(f'(w_+))^{1/2},
\label{eq:cplus}
\eeq
and at $w=w_{-}^s \neq w_+$ satisfying $f(w_{-}^s)=R(w_{-}^s)$ when
\beq
c_{-}=(f'(w_{-}^s))^{1/2},
\label{eq:cminus}
\eeq
Here $c_+$ and $c_{-}$ denote the sound speeds at $w=w_+$ and $w=w_{-}^s$, respectively, and the corresponding Rayleigh lines are shown by blue and green in Fig.~\ref{fig:superkink}. The critical points of \eqref{eq:TW_QC_gen2} rewritten as a first-order system are given by the intersections of $R(w)$ and $f(w)$. A simple analysis shows that for $c_+^2<V^2<c_{-}^2$ there are three such points in the phase plane $(w,w')$: saddle points $(w_+,0)$ and $(w_S,0)$ and a center point $(w_C,0)$. Moreover, for $c_+^2<V^2<V_{SK}^2$, \eqref{eq:TW_QC_gen3} describes a homoclinic trajectory emanating from the saddle point $(w_+,0)$ and corresponding to a tensile solitary wave solution with the background state $w_B=w_+$ at infinity (see the magenta trajectory in Fig.~\ref{fig:pplane}(a) for an example).

As $V^2$ approaches $V^2_{SK}$ from below, the amplitude of the solitary wave (determined by the right hand side of \eqref{eq:TW_QC_gen3}) increases. Its trajectory is passing closer to the saddle point $(w_S,0)$, where $w_S(V)$ approaches $w_{-}$ from above, causing the wave to become more flat and wide in the middle. At $V^2=V_{SK}^2$ the homoclinic orbit reaches the saddle point $(w_{-},0)$, whereupon two heteroclinic trajectories that correspond to superkink solutions form, as shown in Fig.~\ref{fig:pplane}(b). The lower trajectory (marked in red) has the strain $w_{-}$ behind the moving front and $w_{+}$ ahead (a kink), while the upper trajectory (green) has $w_{-}$ ahead and $w_+$ behind (an antikink).

When $V^2$ exceeds $V^2_{SK}$, the heteroclinic orbits are destroyed, and there is another homoclinic trajectory (an example is shown by magenta in Fig.~\ref{fig:pplane}(c)) that emanates from the saddle point $(w_S,0)$ and corresponds to a compressive solitary wave with the background state $w_B=w_S$ that depends on the velocity $V$ of the wave. This trajectory is described by \eqref{eq:TW_QC_gen3} with $w_+$ replaced by $w_S$. Such compressive waves exist for $V_{SK}^2<V^2<c_{-}^2$, where $c_{-}$ is given by \eqref{eq:cminus}. As $V^2 \to c_{-}^2$, $w_S$ approaches $w_{-}^s$, and as $V^2 \to V_{SK}^2$, it tends to $w_{-}$. As $V^2$ approaches $V^2_{SK}$ from above, the width and amplitude of the solitary wave grow, and it becomes more flat in the middle due to its trajectory passing closer to the saddle point $(w_+,0)$. In the sonic limits both tensile and compressive waves delocalize to their background states.

In the above discussion, we chose $w_+$ to be below $w_*$, where $f'(w)$ has a local maximum; recall that this is the strain value associated with the emergence of small-amplitude kink solutions \cite{Iooss00}. The picture is similar when $w_+>w_*$ but in that case the solitary waves leading to the emergence of superkinks as $V^2$ approaches $V^2_{SK}$ from below are compressive, while above $V^2_{SK}$ there are tensile waves. At $w_+=w_*$, we have $c_+^2=V_{SK}^2=c_{-}^2$, and both solitary waves and the superkinks disappear.

To summarize, for a given state $w_+$ ahead, superkinks arise as the limit of solitary wave solutions. As the kink velocity is approached, these solutions grow in amplitude and become wider and more flat in the middle, with the two boundary layers on the left and on the right that approximate monotone kink and antikink solutions. Thus, for velocities just below the kink limit, solitary waves acquire a dipole structure, where a kink and an antikink move in tandem. This will be further illustrated by explicit solutions constructed in the next section.

We remark that broad solitary waves the type we see around $V=V_{SK}$, are sometimes referred to as ``flat-top solitons". They have been seen in a variety of models, including, for example, the recent analysis of the continuum nonlinear equations of  Gardner-type \cite{RosenauOron20,RosenauOron22} and of closely related overdamped discrete oscillator chains \cite{RosenauPikovsky20,RosenauPikovsky21}. In the context of the FPU problem, such solitary wave solutions and the limiting superkinks were first studied for the special case of bilinear interactions in \cite{Gorbushin20,Gorbushin21}, where the  discrete problem could be solved explicitly.

\section{Explicit solutions for the quasicontinuum model}
\label{sec:QCsolns}
The generic scenario described in the previous section holds for any smooth hardening-softening $f(w)$. In the cubic case \eqref{eq:cubic}, we can integrate \eqref{eq:TW_QC_gen3} to obtain explicit solutions that have a simple form. One can show that $f(w)$ in \eqref{eq:cubic} has the symmetry property
\[
f(w)=\dfrac{4b}{3}w_*^2+2w_*-f(2w_*-w),
\]
with $w_*$ given in \eqref{eq:cubic_more}, so if $w(\xi)$ is a traveling wave solution of either discrete or QC problem with velocity $V$, so is $\tilde{w}(\xi)=2w_*-w(\xi)$ with the corresponding adjustment of the conditions at infinity.\\

\noindent {\bf Superkinks.} We begin by constructing a superkink solution. Due to the symmetry it suffices to consider the case $w_{-}>w_{+}$. Note that \eqref{eq:cubic} implies that
\[
\Phi(w_{-})-\Phi(w_{+})-\dfrac{1}{2}(w_{-}-w_{+})(f(w_{+})+f(w_{-}))=-\dfrac{1}{2}(w_{-}-w_{+})^3\left[-\dfrac{a}{2}(w_{+}+w_{-})-\dfrac{b}{3}\right],
\]
so that \eqref{eq:zero_G} yields
\beq
w_{+}+w_{-}=\dfrac{2b}{3|a|}=2w_*.
\label{eq:cubic_cond1}
\eeq
Noting that
\[
f(w_{-})-f(w_{+})=(w_{-}-w_{+})\left[a(w_{+}^2+w_{+}w_{-}+w_{-}^2)+b(w_{+}+w_{-})+1\right]
\]
and using \eqref{eq:cubic_cond1}, we find that \eqref{eq:Rline} yields
\beq
V_{SK}^2=1+\dfrac{b^2}{3|a|}-|a|\left(w_{+}-\dfrac{b}{3|a|}\right)^2,
\label{eq:cubic_cond2}
\eeq
which together with \eqref{eq:cubic_cond1} implies that
\beq
w_\pm=\dfrac{b}{3|a|} \mp \sqrt{\dfrac{b^2}{3a^2}-\dfrac{V_{SK}^2-1}{|a|}}.
\label{eq:wpm_cubic}
\eeq
The expression under the square root must be positive, which yields the upper velocity bound, $V_{SK}^2 < 1+b^2/(3|a|)$. It is reached when $w_{+}=w_{-}=w_*=b/(3|a|)$, the strain value where $f(w)$ changes curvature from convex to concave and the bifurcation point for the superkink solution.

Substituting \eqref{eq:wpm_cubic} in \eqref{eq:TW_QC_gen3} with $f(w)$ given by \eqref{eq:cubic}, we obtain
\beq
\dfrac{V_{SK}^2}{24}(w'(\xi))^2=-\dfrac{a}{4}(w-w_{+})^2(w-w_{-})^2,
\label{eq:TW_QC_cubic}
\eeq
where we recall that $a<0$. Note that $b>0$ then ensures that $w_{+}$ and $w_{-}$ have a positive average   $b/(3|a|)$ , and $f(w)$ monotonically increases in the interval in \eqref{eq:cubic_more} around this average. Requiring that $w_{\pm}$ in \eqref{eq:wpm_cubic} belong to this interval (so that $f'(w_\pm)>0$) gives the lower velocity bound, which together with the upper bound obtained above yields
\beq
\dfrac{2}{3}\left(1+\dfrac{b^2}{3|a|}\right) < V_{SK}^2 < 1+\dfrac{b^2}{3|a|}.
\label{eq:V_bounds_cubic}
\eeq

\begin{figure}
\centering
\includegraphics[width=\textwidth]{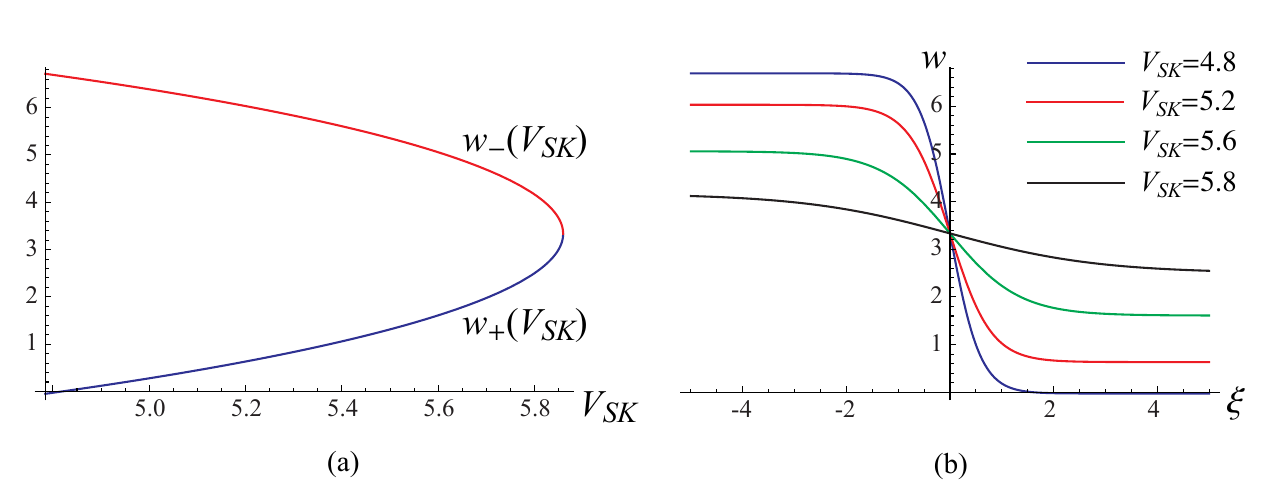}
\caption{\footnotesize (a) The limiting strains $w_\pm$ of a superkink as functions of $V_{SK}$ for the cubic nonlinearity \eqref{eq:cubic} with $a=-1$ and $b=10$. The figure is symmetric about the vertical axis. (b) Strain profiles for the QC model at different velocity values.}
\label{fig:cubic}
\end{figure}

Since $w'(\xi)<0$ along the solution we seek, with $w_{+}<w(\xi)<w_{-}$, \eqref{eq:TW_QC_cubic} yields the separable ordinary differential equation
\[
\dfrac{dw}{d\xi}=-\dfrac{\sqrt{6|a|}}{|V_{SK}|}(w-w_{+})(w_{-}-w),
\]
which is readily solved. Assuming $w(0)=(w_{+}+w_{-})/2=w_*$ (a choice we can make due to translational invariance) and using \eqref{eq:wpm_cubic}, we obtain
\beq
w(\xi)=\dfrac{w_{+}+w_{-}}{2}-\dfrac{w_{-}-w_{+}}{2}\text{tanh}(p\xi), \quad p=\dfrac{\sqrt{6|a|}|w_{-}-w_{+}|}{2|V_{SK}|}=\dfrac{\sqrt{2(b^2-3|a|(V_{SK}^2-1))}}{|V_{SK}|\sqrt{|a|}}.
\label{eq:SKsoln_cubic}
\eeq
As $V_{SK}^2$ increases within the interval in \eqref{eq:V_bounds_cubic}, the values $w_{\pm}$ move toward each other, while the width of the transition front increases. This is illustrated in Fig.~\ref{fig:cubic}.

In the above construction, we assumed that $w_{-}>w_{+}$. Due to the symmetry mentioned above, solutions with $w_{+}>w_{-}$ have the same form \eqref{eq:SKsoln_cubic} but $\mp$ in the right hand side of \eqref{eq:wpm_cubic} becomes $\pm$.\\

\noindent {\bf Solitary waves.} We now consider solitary wave solutions associated with a superkink that has the state $w_+ \neq w_*$ ahead and velocity $V_{SK}$ given by \eqref{eq:cubic_cond2}. Recall from Sec.~\ref{sec:QC} that such waves have velocity $V$ such that either $c_{+}^2<V^2<V_{SK}^2$ or $V_{SK}^2<V^2<c_{-}^2$, where $c_+$ and $c_{-}$ are defined in \eqref{eq:cplus} and \eqref{eq:cminus}. Recall also that the waves have the background state
\beq
w_B=\begin{cases} w_+, & c_+^2<V^2<V_{SK}^2,\\
                  w_S(V), & V_{SK}^2<V^2<c_{-}^2,
    \end{cases}
\label{eq:wB}
\eeq
where $w_S \neq w_+$ is such that $f(w_S)-f(w_+)=V^2(w_S-w_+)$ and $V^2>f'(w_S)$, so that $(w_S,0)$ corresponds to a saddle point in the phase plane for \eqref{eq:TW_QC_gen2}.

\begin{figure}
\centering
\includegraphics[width=0.5\textwidth]{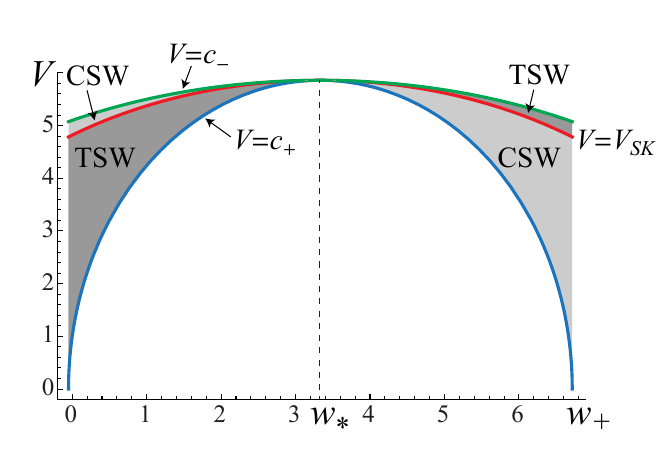}
\caption{\footnotesize Velocities of the solitary waves as a function of the state $w_+$ ahead of the associated superkink solution. The red curve corresponds to the superkink limit, while the blue and green curves indicate the sonic limits. Dark shaded regions correspond to tensile solitary waves (TSW), and light shaded regions to compressive solitary waves (CSW). The diagram is symmetric about the horizontal axis. Here $f(w)$ is given by \eqref{eq:cubic} with $a=-1$, $b=10$.}
\label{fig:SWvelocities_cubic}
\end{figure}

In the cubic case \eqref{eq:cubic} we use
\[
f(w_S)-f(w_{+})=(w_S-w_{+})\left[a(w_+^2+w_{+}w_S+w_S^2)+b(w_{+}+w_S)+1\right]
\]
to obtain
\beq
w_S=\dfrac{1}{2|a|}[b-|a|w_+ \mp \sqrt{b^2-3a^2w_+^2+2|a|bw_{+}-4|a|(V^2-1)}],
\label{eq:wS}
\eeq
with the minus sign if $w_+>w_*$ and plus sign if $w_+<w_*$. To find $c_{-}$ in \eqref{eq:cminus}, we recall that
$f(w_{-}^s)-f(w_+)=f'(w_{-}^2)(w_{-}^s-w_+)$. Substituting \eqref{eq:cubic}, we obtain, after some algebra,
\[
w_{-}^s=\dfrac{1}{2|a|}(b-|a| w_{+}),
\]
and hence $c_{-}^2=f'(w_{-}^s)=3a(w_{-}^s)^2+2bw_{-}^s+1$ yields
\beq
c_{-}=\sqrt{\dfrac{1}{4}(b-|a|w_+)\left(3w_{+}+\dfrac{b}{|a|}\right)+1}.
\label{eq:cminus_cubic}
\eeq
Meanwhile, $c_{+}=(f'(w_{+}))^{1/2}=(3a(w_{+})^2+2bw_{+}+1)^{1/2}$. For $w_+ \neq w_*$ we have $c_+^2<V_{SK}^2<c_{-}^2$, as illustrated in Fig.~\ref{fig:SWvelocities_cubic}. At $w_+=w_*$ the three velocities coincide: $c_+^2=V_{SK}^2=c_{-}^2=1-b^2/(3a)$.

For solitary waves \eqref{eq:TW_QC_gen3} with $f(w)$ given by \eqref{eq:cubic} has the form
\beq
\dfrac{V^2}{24}(w'(\xi))^2=-\dfrac{a}{4}(w-w_B)^2(w_T-w)(w_M-w),
\label{eq:SW_QC_cubic}
\eeq
where the equilibrium points $w_T$ and $w_M$ are given by
\beq
w_{T,M}=\dfrac{2b}{3|a|}-w_B\pm\sqrt{4\left(w_B-\dfrac{b}{3|a|}\right)^2-\dfrac{2}{|a|}(V^2-1)-6w_B^2+\dfrac{4b}{|a|}w_B},
\label{eq:w_T_M}
\eeq
with plus sign in front of the square root for $w_T$ and minus for $w_M$ when $w_+<w_*$, $c_+^2<V^2<V_{SK}^2$ or $w_+>w_*$, $V_{SK}^2<V^2<c_{-}^2$ and vice versa when $w_+>w_*$, $c_+^2<V^2<V_{SK}^2$ or $w_+<w_*$, $V_{SK}^2<V^2<c_{-}^2$, and we recall that $a<0$. Solving \eqref{eq:SW_QC_cubic}, we obtain the solitary wave
\beq
w(\xi)=w_B+\dfrac{2(w_M-w_B)(w_T-w_B)}
{w_M-2w_B+w_T+(w_T-w_M)\cosh(\gamma\xi)},\;\;\gamma=\dfrac{\sqrt{6|a|}}{|V|}\sqrt{(w_M-w_B)(w_T-w_B)},
\label{eq:SWsoln_cubic}
\eeq
where we recall \eqref{eq:wB} and \eqref{eq:wS}. As shown in Fig.~\ref{fig:SWvelocities_cubic}, for $w_+<w_*$, the solitary waves are tensile when $c_+^2<V^2<V_{SK}^2$ and compressive when $V_{SK}^2<V^2<c_{-}^2$, and the opposite is true for $w_+>w_*$.

\begin{figure}[h]
\centering
\includegraphics[width=\textwidth]{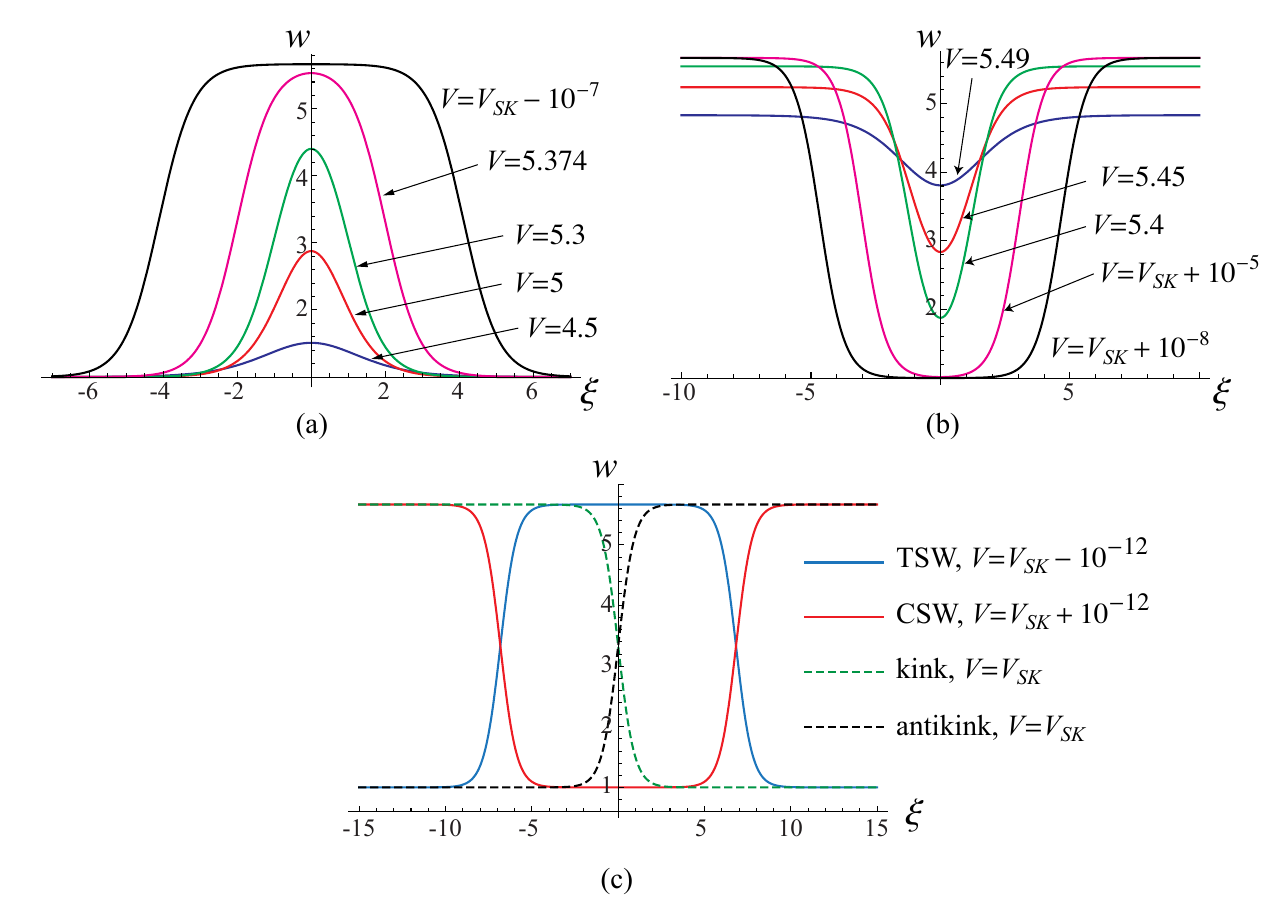}
\caption{\footnotesize (a) Tensile solitary waves (TSW) in the QC model with cubic nonlinearity \eqref{eq:cubic} at $c_+<V<V_{SK}$; (b) compressive solitary wabes (CSW) at $V_{SK}<V<c_{-}$; (c) solitary waves just below and just above $V_{SK}$, shown together with kink and antikink fronts. Here $w_+=1$, $a=-1$, $b=10$, yielding $V_{SK}=2\sqrt{65}/3$, $w_{-}=17/3$, $c_{+}=3\sqrt{2}$, $c_-=11/2$.}
\label{fig:SWs_cubic}
\end{figure}
The solution \eqref{eq:SWsoln_cubic} satisfies $w(0)=w_M$ and $w(\xi) \to w_B$ as $\xi \to \pm \infty$. The amplitude of the solitary wave is thus given by
\beq
w_{amp}^{QC}=|w_M-w_B|
\label{eq:SWQC_amp}
\eeq
Note that the amplitude tends to zero (solution delocalizes to the constant strain $w_B$) as $V$ tends to the corresponding sonic limit. As the superkink velocity limit, $V_{SK}$, is approached, $w_B \to w_{-}$, $w_{T,M} \to w_{+}$ for $V_{SK}^2<V^2<c_{-}^2$. Meanwhile, for $c_+^2<V^2<V_{SK}^2$ we have $w_B=w_{+}$, and $w_{T,M} \to w_{-}$ in the superkink limit. Thus
\[
w_{amp}^{QC} \to |w_{-}-w_{+}|,
\]
where $w_{-}$ is the strain behind the superkink front corresponding to $w_+$, and we recall \eqref{eq:cubic_cond1}.
Note also that in this limit $\gamma$ in \eqref{eq:SWsoln_cubic} tends to $2 p$, with $p$ defined in \eqref{eq:SKsoln_cubic}. Thus, as the superkink velocity is approached, the solitary wave \eqref{eq:SWsoln_cubic} becomes wider, with the two boundary layers on the left and on the right approaching the corresponding superkink solutions, and the strain in between tending to the constant value given by $w_{-}$ for $c_+^2<V^2<V_{SK}^2$ and $w_+$ for $V_{SK}^2<V^2<c_{-}^2$. Just below and just above the limit, solitary waves have the structure of a kink-antikink bundle. This is illustrated in Fig.~\ref{fig:SWs_cubic}.

\begin{figure}
\centering
\includegraphics[width=\textwidth]{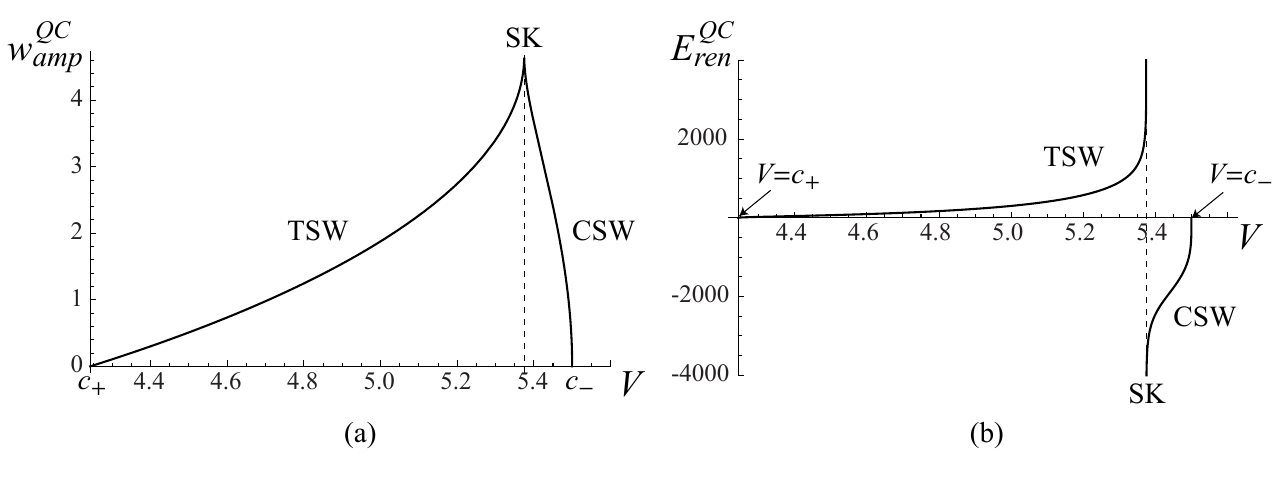}
\caption{\footnotesize (a) Amplitude \eqref{eq:SWQC_amp} and (b) renormalized energy \eqref{eq:Eren_QC} as a function of velocity $V$ for solitary waves in QC model at $w_+=1$. Here $a=-1$, $b=10$, and the superkink (SK) limit $V_{SK}=2\sqrt{65}/3$ is marked by the dashed vertical line. The sonic limits are $c_{+}=3\sqrt{2}$ and $c_-=11/2$.}
\label{fig:SWamp_energy_QCcubic}
\end{figure}
Since the energy of the waves with nonzero background is infinite, we renormalize it by subtracting the energy of the background:
\beq
E^{QC}_{ren}(V)=\int_{-\infty}^{\infty} \left\{\dfrac{1}{2}V^2 w^2(\xi)+\dfrac{1}{24}V^2 (w'(\xi))^2+\Phi(w(\xi))-\Phi(w_B)-\dfrac{1}{2}V^2 w_B^2\right\}d\xi,
\label{eq:Eren_QC}
\eeq
where we used the fact that for a traveling wave solution with the strain $w(\xi)=w(x-Vt)$ the particle velocity is $v(\xi)=-V w(\xi)$.
Fig.~\ref{fig:SWamp_energy_QCcubic} shows the typical dependence of amplitude and renormalized energy of the waves on their velocity. As discussed above, in the superkink limit (marked by the dashed vertical line) the amplitude reaches the finite value $|w_{-}-w_{+}|$ (a corner in Fig.~\ref{fig:SWamp_energy_QCcubic}(a)), while the renormalized energy diverges near the limit (see Fig.~\ref{fig:SWamp_energy_QCcubic}(b)) because the two superkinks forming such solitary waves undergo an unlimited separation.

Thus, we can see already at the QC level that the superkink and solitary wave solutions form a single family, with a singular superkink limit embedded in the continuum range of solitary wave velocities.

\section{Traveling wave solutions of the discrete problem}
\label{sec:Dsolns}
Having explored the relation between superkinks and solitary waves on the QC level, we now consider the corresponding traveling wave solutions of the discrete problem \eqref{eq:FPUstrain}. Due to the symmetry of the problem with respect to velocity $V$, it suffices to obtain solutions with $V>0$.\\

\noindent {\bf Superkinks.} To compute the superkink solutions, we follow the approach in \cite{Aubry09,Vainchtein20,James21} and observe that by virtue of the traveling wave ansatz \eqref{eq:TWansatz} such solutions are necessarily periodic modulo shift by one lattice space,
\beq
w_{n+1}(t+T)=w_n(t), \quad T=1/V,
\label{eq:period}
\eeq
and thus can be cast as fixed points of the nonlinear map
\begin{equation}
\label{eq:nonlin_map}
\left[\begin{array}{c}
  \{w_{n+1}(T)\} \\ \{\dot w_{n+1}(T)\} \\  \end{array}\right]
  =\mathcal{N}\left(
  \left[\begin{array}{c}
  \{w_{n}(0)\} \\ \{\dot w_{n}(0)\} \\  \end{array}\right]\right)
\end{equation}
defined by integration of the governing equations \eqref{eq:FPUstrain} over one period followed by a shift of indices.
To obtain the traveling waves, we follow an approach used in computing discrete breathers \cite{MarinAubry96} and employ the fixed point method.
For a large even number $N$ (we used $N=500$ in a typical computation) and given $T=1/V$, we perform the Newton-Raphson iterations   with numerically computed finite-difference Jacobian to solve
\beq
\begin{split}
&w_{n+1}(T)=w_n(0), \quad n=-N/2,\dots,N/2-1, \\
&\dot{w}_{n+1}(T)=\dot{w}_n(0), \quad n=-N/2,\dots,N/2-2, \quad w_1(T)=w_*
\end{split}
\label{eq:nonlin_system}
\eeq
for $\{w_n(0),\dot{w}_n(0)\}$, $n=-N/2,\dots N/2-1$.
To obtain $w_n(T)$ and $\dot{w}_n(T)$ for given $w_n(0)$ and $\dot{w}_n(0)$ at each iteration, we integrate \eqref{eq:FPUstrain} over one period using the Dormand-Prince algorithm (Matlab's ode45 routine) with boundary conditions
\beq
w_{-N/2-1}(t)=w_{-}, \quad w_{N/2}(t)=w_{+},
\label{eq:fixedBCs}
\eeq
where $w_{\pm}$ are found from \eqref{eq:Rline}, \eqref{eq:zero_G}. The last equation in \eqref{eq:nonlin_system} represents a pinning condition. Due to translational invariance of solutions of \eqref{eq:TW_discrete}, such condition is necessary to select a unique traveling wave solution. The one we select facilitates the comparison with superkink solutions $w_{QC}(\xi)$ of the QC model in \eqref{eq:SKsoln_cubic}, which are also used to obtain an initial guess for the Newton-Raphson procedure and parameter continuation. Recall that these solutions satisfy $w_{QC}(0)=w_*$, so that $w_0(0)=w_1(T)=w_{QC}(0)$ and thus the traveling wave $w(\xi)$ for the discrete problem satisfies $w(0)=w_{QC}(0)$. We drop the equation for $\dot{w}_{N/2}(T)$ in \eqref{eq:nonlin_system} in order to obtain a system of $2N$ nonlinear equations for $2N$ unknowns while prescribing the pinning condition. We have verified that the omitted equation is automatically satisfied up to the order of $10^{-13}$ at most in the computed solutions, due to the large value of $N$.

\begin{figure}
\centering
\includegraphics[width=0.5\textwidth]{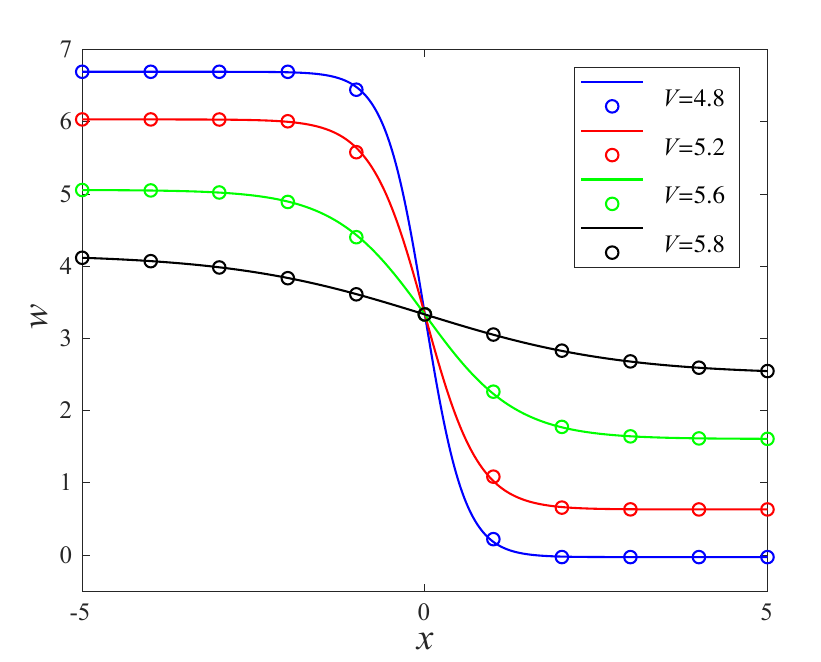}
\caption{\footnotesize Superkink solutions $w_n(0)=w(n)$ of the discrete problem \eqref{eq:TW_discrete} (circles) with cubic nonlinearity \eqref{eq:cubic} and the corresponding solutions $w(x)$ for the QC model \eqref{eq:TW_QC_gen3} (solid curves) evaluated at $t=0$. Here $a=-1$, $b=10$.}
\label{fig:TWcubic}
\end{figure}

The computed superkink profiles $w_n(0)=w(n)$ for are shown in Fig.~\ref{fig:TWcubic}, together with the corresponding profiles $w(x)$ obtained from the exact solutions \eqref{eq:SKsoln_cubic} of the QC model. One can see that these solutions are very close, with barely visible difference in the transition layer.\\

\noindent {\bf Solitary waves.} To compute the solitary wave solutions for given $w_+$ and velocity $V$ in the intervals $(c_+,V_{SK})$ and $(V_{SK},c_{-})$, we use the same approach as for the superkinks. In this case the prescribed pinning condition is $\dot{w}_1(T)=0$, to ensure that the maximum of a tensile solitary wave (or the minimum of a compressive one) is at $n=0$ when $t=0$, and the boundary conditions are $w_{-N/2-1}(t)=w_{N/2}(t)=w_B$, where we recall \eqref{eq:wB}.

\begin{figure}
\centering
\includegraphics[width=\textwidth]{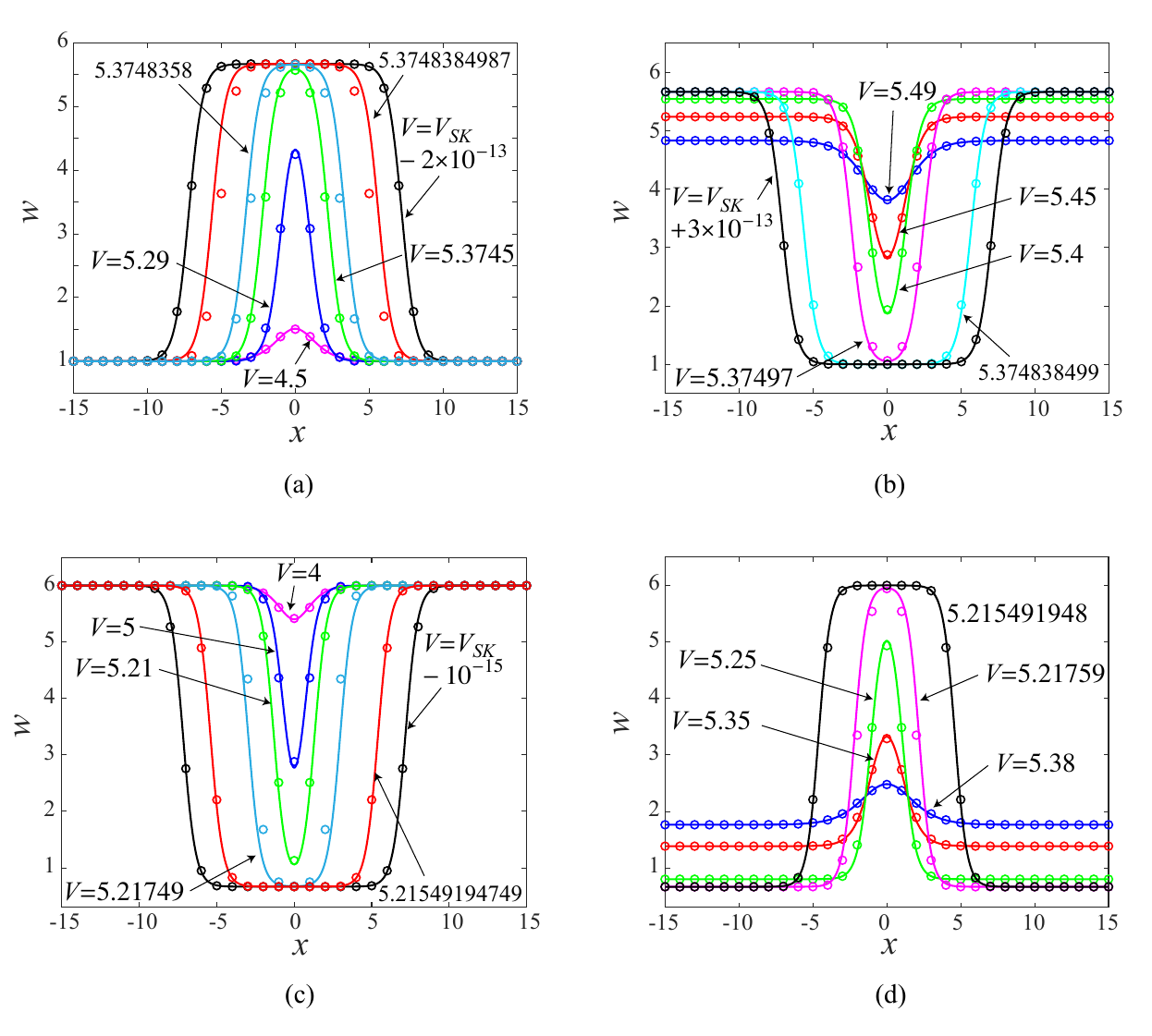}
\caption{\footnotesize Solitary wave solutions $w_n(0)=w(n)$ of the discrete problem \eqref{eq:TW_discrete} (circles) with cubic nonlinearity \eqref{eq:cubic} and the corresponding solutions $w(x)$ for the QC model \eqref{eq:TW_QC_gen3} (solid curves) evaluated at $t=0$. The top two panels show (a) tensile waves below the superkink limit and (b) compressive waves above it at $w_+<w_*$. The two bottom panels show (c) compressive waves below the superkink limit and (d) tensile waves above it at $w_+>w_*$. Here $a=-1$, $b=10$, and the values of $w_+$ and the corresponding superkink velocity $V_{SK}$ and sonic speeds $c_{\pm}$ are $w_+=1$, $V_{SK}=2\sqrt{65}/3$, $c_{+}=3\sqrt{2}$, $c_-=11/2$ in (a), (b) and $w_+=6$, $V_{SK}=7\sqrt{5}/3$, $c_{+}=\sqrt{13}$, $c_{-}=\sqrt{29}$ in (c), (d).}
\label{fig:SWs_cubic_DvsQC}
\end{figure}

\begin{figure}
\centering
\includegraphics[width=\textwidth]{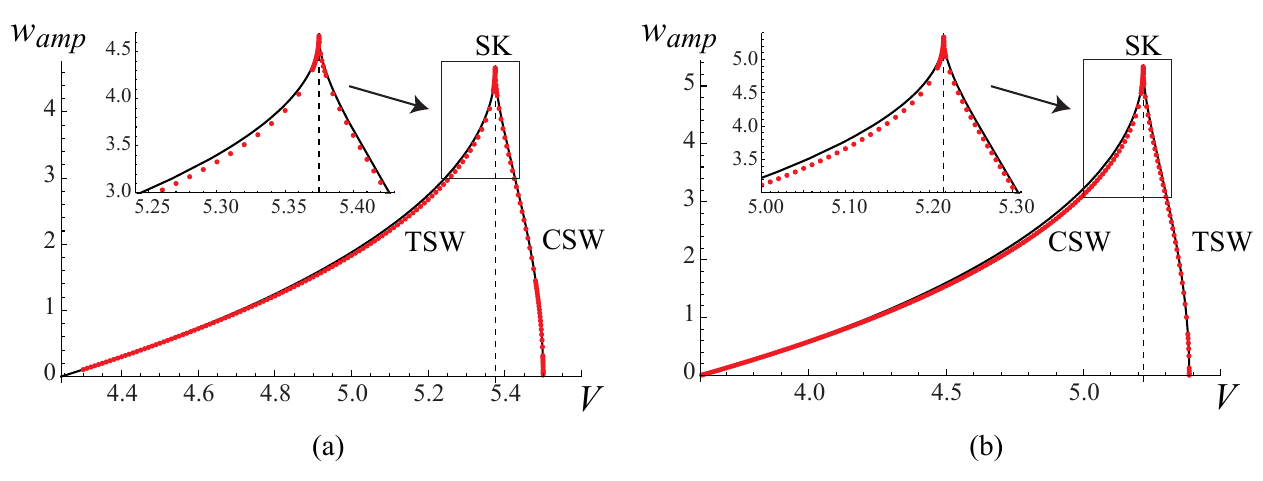}
\caption{\footnotesize (a) Amplitude $w_{amp}=|w(0)-w_+|$ as a function of velocity $V$ for solitary wave solutions of the discrete problem \eqref{eq:TW_discrete} (dots) with cubic nonlinearity \eqref{eq:cubic} and the corresponding solutions for the QC model \eqref{eq:TW_QC_gen3} (solid curves): (a) $w_+<w_*$; (b) $w_+>w_*$. The superkink limit (SK) is marked by the dashed vertical line, and tensile and compressive waves are marked by TSW and CSW, respectively. Here $a=-1$, $b=10$, and the values of $w_+$ and the corresponding superkink velocity $V_{SK}$ and sonic speeds $c_{\pm}$ are $w_+=1$, $V_{SK}=2\sqrt{65}/3$, $c_{+}=3\sqrt{2}$, $c_-=11/2$ in (a) and $w_+=6$, $V_{SK}=7\sqrt{5}/3$, $c_{+}=\sqrt{13}$, $c_{-}=\sqrt{29}$ in (b). Insets zoom in inside the rectangles.}
\label{fig:SWamp_cubic_DvsQC}
\end{figure}

\begin{figure}
\centering
\includegraphics[width=\textwidth]{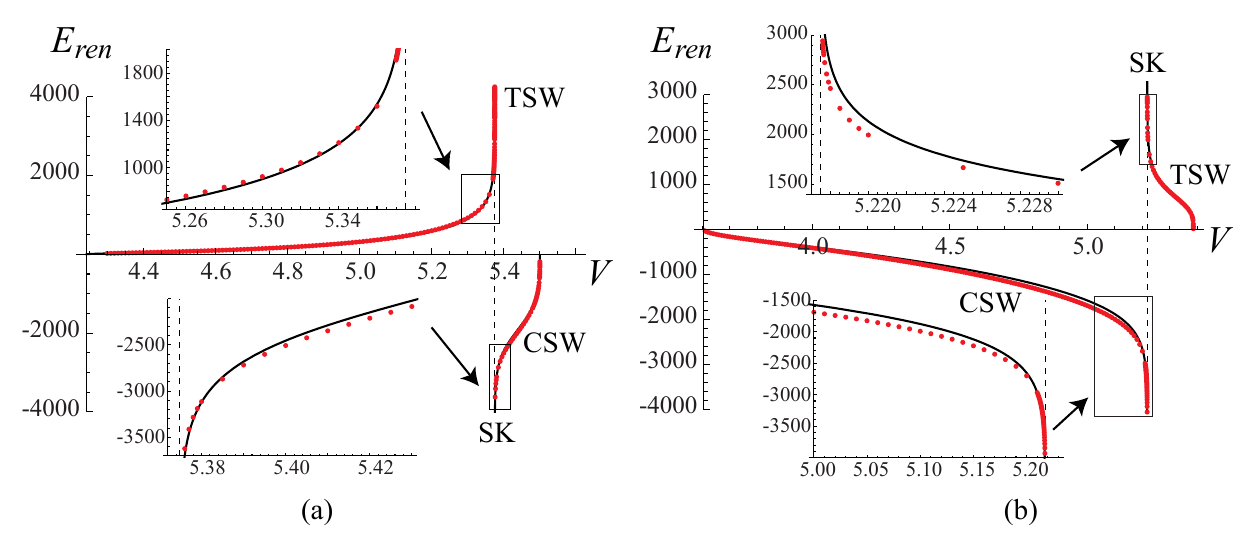}
\caption{\footnotesize Renormalized energy $E_{ren}$ given by \eqref{eq:Eren_D} as a function of velocity $V$ for solitary wave solutions of the discrete problem \eqref{eq:TW_discrete} (dots) with cubic nonlinearity \eqref{eq:cubic} and the corresponding energy \eqref{eq:Eren_QC} for the QC model \eqref{eq:TW_QC_gen3} (solid curves): (a) $w_+<w_*$; (b) $w_+>w_*$. The superkink limit (SK) is marked by the dashed vertical line, and tensile and compressive waves are marked by TSW and CSW, respectively. Here $a=-1$, $b=10$, and the values of $w_+$ and the corresponding superkink velocity $V_{SK}$ and sonic speeds $c_{\pm}$ are $w_+=1$, $V_{SK}=2\sqrt{65}/3$, $c_{+}=3\sqrt{2}$, $c_-=11/2$ in (a) and $w_+=6$, $V_{SK}=7\sqrt{5}/3$, $c_{+}=\sqrt{13}$, $c_{-}=\sqrt{29}$ in (b). Insets zoom in inside the rectangles.}
\label{fig:SWenergy_cubic_DvsQC}
\end{figure}

The resulting strain profiles are shown in Fig.~\ref{fig:SWs_cubic_DvsQC} together with their QC counterparts \eqref{eq:SWsoln_cubic}. For further comparison of solitary waves in the discrete and QC models, we show the corresponding amplitude-velocity plots in Fig.~\ref{fig:SWamp_cubic_DvsQC} and energy-velocity plots in Fig.~\ref{fig:SWenergy_cubic_DvsQC}.
In the latter, we compare the renormalized energy \eqref{eq:Eren_QC} for the QC model with the corresponding values
\beq
E^D_{ren}(V)=\sum_n \left\{\dfrac{1}{2}v_n^2+\dfrac{1}{2}\left(\Phi(w_n)+\Phi(w_{n+1})\right)-\Phi(w_B)-\dfrac{1}{2}V^2 w_B^2\right\}
\label{eq:Eren_D}
\eeq
in the discrete case, where we recall that $v_n$ are the particle velocities, and all values are evaluated at $t=0$ due to the energy conservation.

One can see that in the case of solitary waves some discrepancy between solutions of the discrete and QC problems is visible away from the sonic and superkink limits. However, the QC model still provides a very good quantitative approximation of the entire solution family. Importantly, it captures the singular nature of the superkink limit as well as near-sonic regimes exceptionally well.

\section{Stability of superkinks and solitary waves}
\label{sec:stab}
To investigate the linear stability of the obtained traveling wave solutions in the problem with cubic nonlinearity \eqref{eq:cubic}, we follow the approach in \cite{Cuevas17,Xu18,Vainchtein20} and use Floquet analysis that exploits periodicity-modulo-shift \eqref{eq:period} of the traveling wave solutions. Substituting $w_n(t)=\hat{w}_n(t)+\epsilon y_n(t)$ into \eqref{eq:FPUstrain}, where $\hat{w}_n(t)=w(n-Vt)$ is the traveling wave solution, and considering $O(\epsilon)$ terms, we obtain the governing equations for the linearized problem:
\beq
\ddot{y}_n=f'(\hat{w}_{n+1})y_{n+1}-2f'(\hat{w}_n)y_n+f'(\hat{w}_{n-1})y_{n-1}.
\label{eq:linear}
\eeq
The Floquet multipliers $\mu$ for this problem are the eigenvalues of the monodromy matrix $\mathcal{M}$ defined by
\begin{equation}
\label{eq:linear_map}
\left[\begin{array}{c}
  \{y_{n+1}(T)\} \\ \{\dot y_{n+1}(T)\} \\  \end{array}\right]
  =\mathcal{M}
  \left[\begin{array}{c}
  \{y_{n}(0)\} \\ \{\dot y_{n}(0)\} \\  \end{array}\right].
\end{equation}
To obtain $\mathcal{M}$, we compute the fundamental solution matrix $\mathbf{\Psi}(T)$, which maps $[\{y_{n}(0)\}, \{\dot y_{n}(0)\}]^T$ onto $[\{y_{n}(T)\}, \{\dot y_{n}(T)\}]^T$, $n=-N/2,\dots,N/2-1$, for the first-order linear system equivalent to \eqref{eq:linear}. We use periodic boundary conditions $y_{N/2}(t)=y_{-N/2}(t)$, $y_{-N/2-1}(t)=y_{N/2-1}(t)$, which is justified by the fact that for both solitary waves and superkinks in the problem with cubic nonlinearity \eqref{eq:cubic} the values $f'(\hat{w}_n)$ at the two ends of a large chain rapidly approach the same constant value. We then shift the rows of $\mathbf{\Psi}(T)$ up by one row in the two parts of the matrix corresponding to $y_n$ and $\dot{y}_n$, respectively, with the last row in each part replaced by the first, obtaining $\mathcal{M}$ in \eqref{eq:linear_map}.

The Floquet multipliers are related to the eigenvalues $\lambda$ of the linearization operator via $\mu=e^{\lambda/V}$, and thus $|\mu|>1$ ($\text{Re}(\lambda)>0$) corresponds to instability. The Hamiltonian nature of the problem means that there are quadruples of non-real Floquet multipliers, i.e., if $\mu$ is a multiplier, than so are $\bar{\mu}$, $1/\mu$ and $1/\bar{\mu}$, while the real multipliers come in pairs $\mu$ and $1/\mu$. Linear stability thus requires that all Floquet multipliers lie on the unit circle: $|\mu|=1$.

To further explore stability of the waves, we complement the Floquet analysis by direct numerical simulations.

\subsection{Stability of superkinks}
\begin{figure}
\centering
\includegraphics[width=0.5\textwidth]{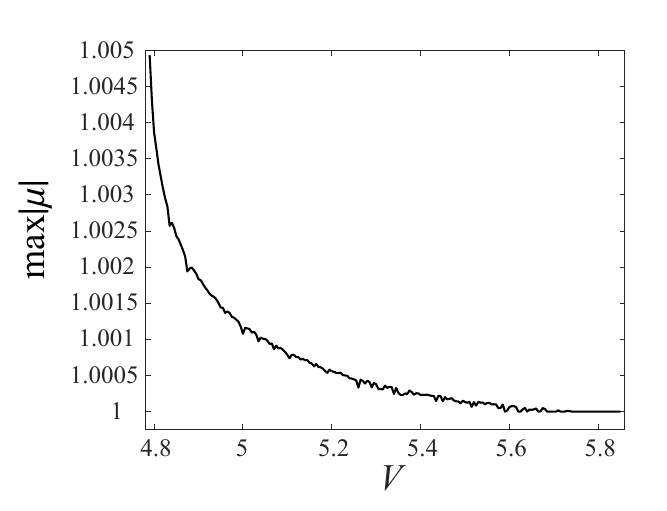}
\caption{\footnotesize Maximum modulus of Floquet multipliers for superkink solutions in the discrete problem with cubic nonlinearity \eqref{eq:cubic}. Here $a=-1$, $b=10$.}
\label{fig:floquet_SK}
\end{figure}
\begin{figure}
\centering
\includegraphics[width=\textwidth]{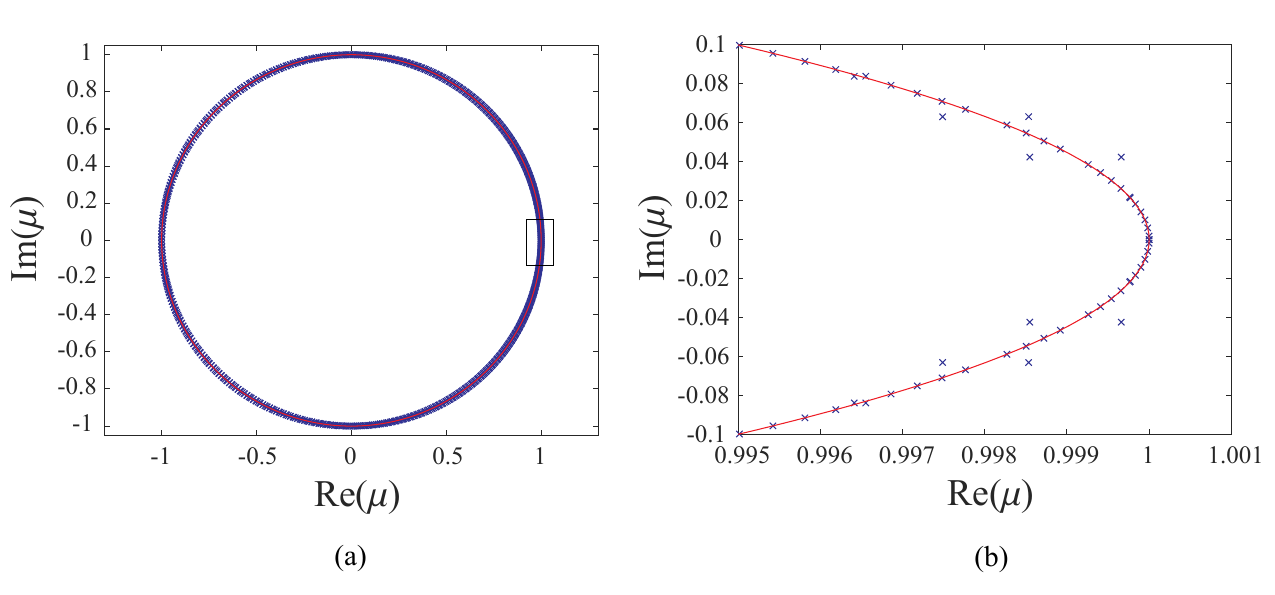}
\caption{\footnotesize (a) Floquet multipliers $\mu$ (blue crosses) for the superkink solution of the discrete problem with cubic nonlinearity \eqref{eq:cubic} at $V=5.2$. (b) Enlarged version of the region inside the rectangle in (a) showing multipliers with $|\mu|>1$ that correspond to mild oscillatory instabilities. The unit circle is shown in red. Here $a=-1$, $b=10$.}
\label{fig:Floquet_example}
\end{figure}
Consider the superkink solutions in the case of cubic nonlinearity \eqref{eq:cubic} with $a=-1$, $b=10$, which are shown in Fig.~\ref{fig:TWcubic}. In this case the velocity range \eqref{eq:V_bounds_cubic} for the superkink traveling waves is $4.78423<V<5.85947$. The results of our Floquet computation with $N=500$, shown in Fig.~\ref{fig:floquet_SK}, indicate that for $V \geq 5.74$ within this range the maximum modulus of the Floquet multipliers exceeds $1$ by $O(10^{-8})$ at most, and thus the corresponding solutions may be considered linearly stable within the accuracy of numerical computation. Below this threshold, quartets of complex multipliers corresponding to oscillatory instability modes emerge from the unit circle for some velocities, as illustrated in Fig.~\ref{fig:Floquet_example}, and rejoin it for others. However, the associated instabilities remain small in magnitude ($|\mu|<1.0001$) for $V \geq 5.57$, and the maximum modulus, which exhibits an overall growth as velocity is decreased toward the lower bound, stays below $1.005$ over the entire velocity range. It should be noted that the Floquet multipliers associated with the oscillatory instabilities depend on the chain size; in particular, their magnitudes decrease as the chain size is increased. This suggests that similar to the case of discrete breathers \cite{Marin98} and solitary waves \cite{Xu18}, these mild instabilities are a spurious artifact of the finite chain size and disappear as $N$ tends to infinity.

\begin{figure}
\centering
\includegraphics[width=\textwidth]{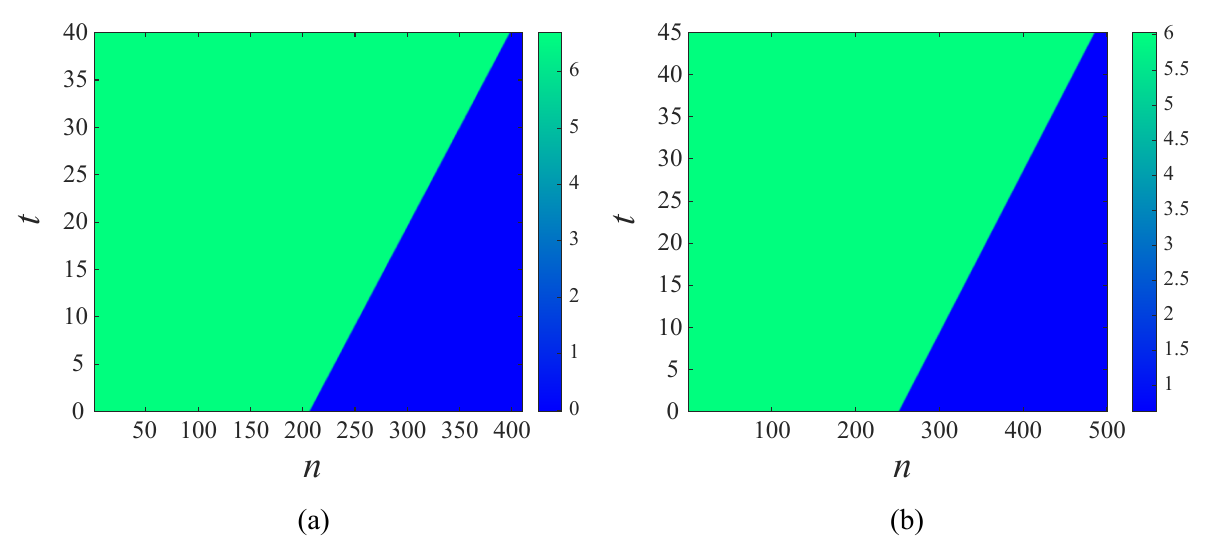}
\caption{\footnotesize Space-time strain evolution initiated by computed superkink solutions with velocities
(a) $V=4.8$; (b) $V=5.2$. Here $a=-1$, $b=10$.}
\label{fig:stabSK}
\end{figure}

In addition to the Floquet analysis, we tested stability of the superkinks by conducting numerical simulations of \eqref{eq:FPUstrain} on a finite chain using the Dormand-Prince algorithm. In the first set of simulations, we extracted initial conditions from the computed superkink solutions, i.e., set $w_n(0)=\hat{w}_n(0)$ and $\dot{w}_n(0)=\dot{\hat{w}}_n(0)$, and used the corresponding fixed boundary conditions \eqref{eq:fixedBCs}. These simulations resulted in steady propagation of the traveling wave with velocity that remained within $O(10^{-8})$ or less from the prescribed value for the entire range of velocities, suggesting that the traveling waves are at least long-lived and likely stable.
Representative examples are shown in Fig.~\ref{fig:stabSK}.

The second set of simulations was conducted on a chain with $L$ particles using free-end boundary conditions and Riemann initial data
\beq
w_n(0)=\begin{cases}w^{l}, & 1 \leq n \leq L/2\\
                    w^{r}, & L/2+1 \leq n \leq L
       \end{cases},
\qquad \dot{w}_n(0)=0, \quad n=1,\dots,L,
\label{eq:Riemann_data}
\eeq
where the left strain satisfies $w_*<w^{l}<(b+\sqrt{b^2+3|a|})/(3|a|)$, in accordance with the bounds for the limiting strain behind a superkink, and we set the right strain to zero: $w^{r}=0$. The strain values obtained in the simulations remained within the region where $f'(w)>0$. The size $L$ of the chain was chosen sufficiently large to avoid any boundary effects.

\begin{figure}
\centering
\includegraphics[width=0.95\textwidth]{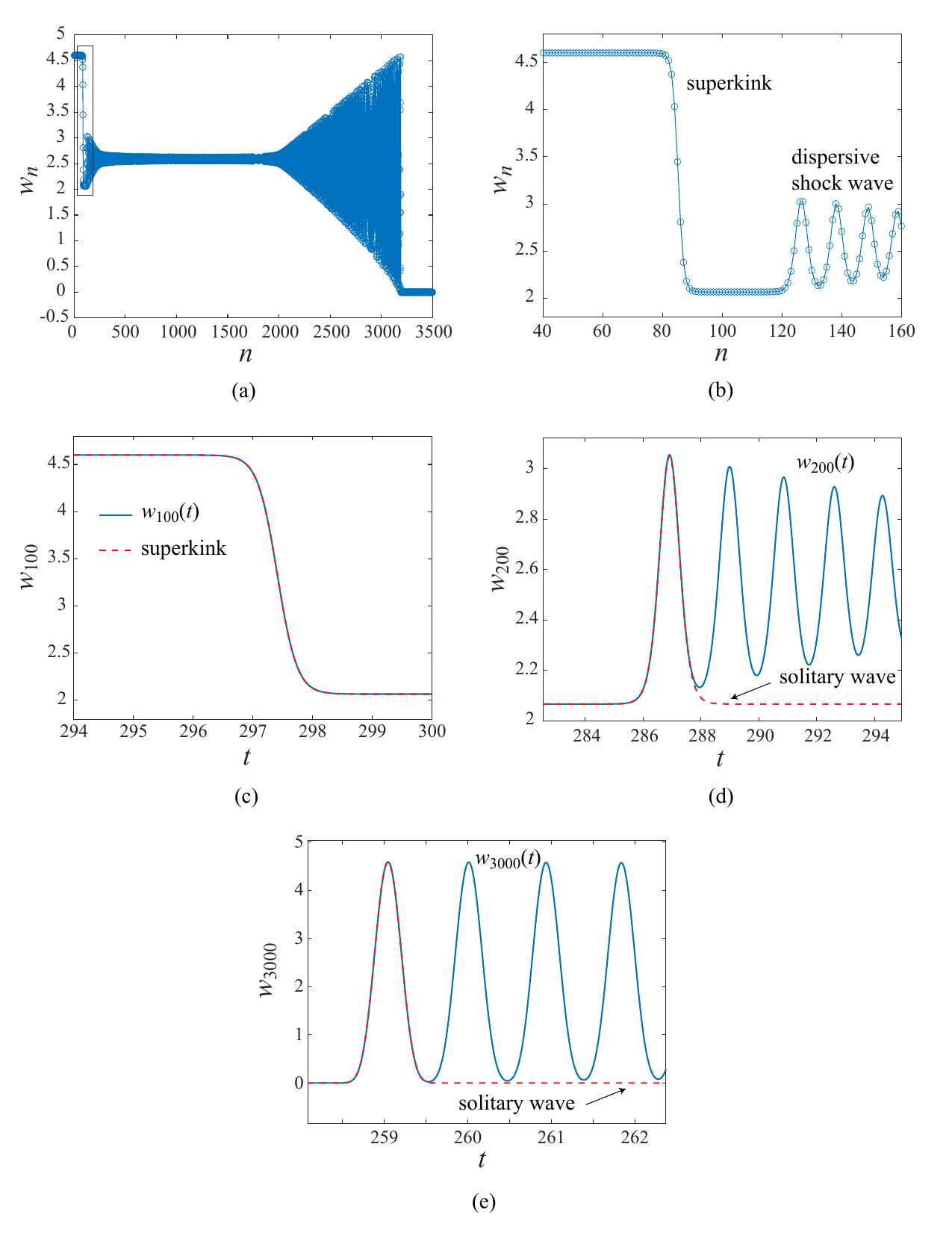}
\caption{\footnotesize The results of simulations with Riemann initial data \eqref{eq:Riemann_data} with $w^{l}=4.6$, $w^{r}=0$, $L=3600$ and cubic nonlinearity \eqref{eq:cubic}: (a) strain profile at $t=300$; (b) enlarged version of the region inside the rectangle in (a); (c) superkink with $V=-5.7209$ (red dashed curve) and $w_{100}(t)$ (blue curve); (d) solitary wave with $V=-5.6159$, $w_B=2.0667$ (red dashed curve) and $w_{200}(t)$ (blue curve); (e) solitary wave with $V=4.6404$, $w_B=0$ (red dashed curve) and $w_{3000}(t)$ (blue curve). Here $a=-1$, $b=10$.}
\label{fig:Riemann1}
\end{figure}
As a representative example, we show the results for $w^{l}=4.6$ in Fig.~\ref{fig:Riemann1}. One can see that the initial data leads to formation of two non-stationary (spreading) dispersive shock waves propagating in opposite directions and a superkink that travels to the left ahead of the corresponding dispersive shock wave (DSW)
\cite{HerrmannRademacher10a,el2016dispersive,kamchatnov2019dispersive,purohit2022dissipation,chong2022dispersive}.
The numerically measured velocity of the superkink, $V=-5.7209$, coincides up to $O(10^{-8})$ with the value associated with the prescribed $w^{l}$; see also the comparison of the (appropriately shifted) computed superkink solution and $w_{100}(t)$ in panel (c).
Formation of the superkink front from generic initial conditions indicates its effective stability, in agreement with the results of the Floquet analysis. The velocity of the leading edge of the weak DSW moving to the left behind the superkink is $V_{DSW_l} = -5.6159$, while the strong DSW moving to the right propagates with $V_{DSW_r} = 4.6404$. As shown in panels (d) and (e), their leading edges are well approximated by computed solitary wave solutions with the corresponding velocities and background strains.

\subsection{Stability of solitary waves}
\begin{figure}
\centering
\includegraphics[width=\textwidth]{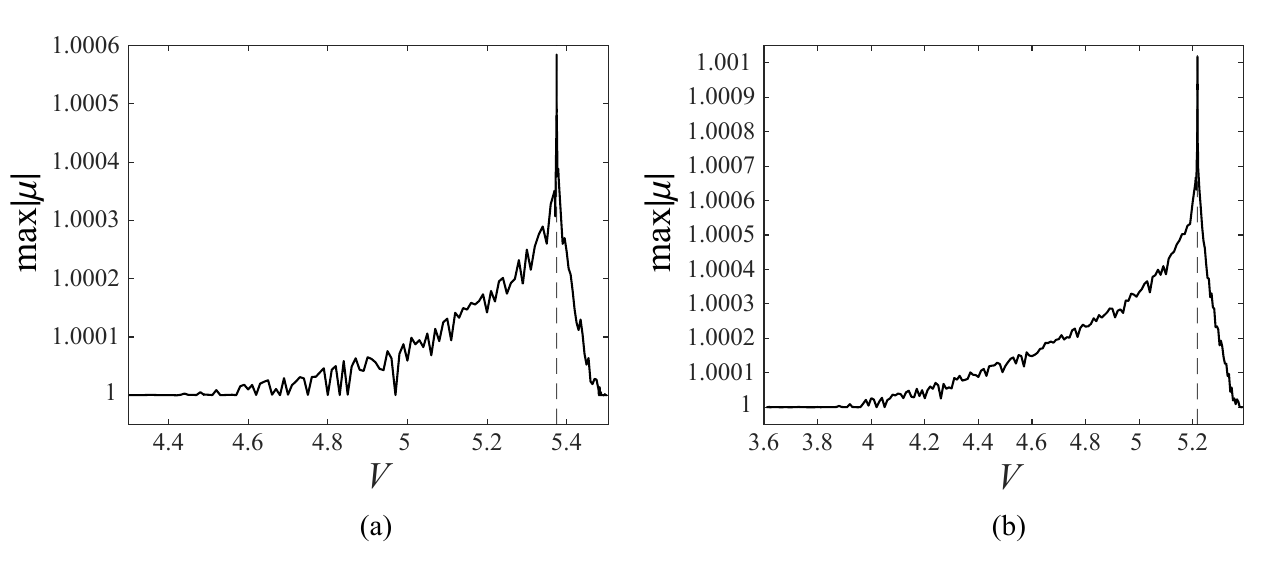}
\caption{\footnotesize Maximum modulus of Floquet multipliers for solitary waves in the discrete problem with cubic nonlinearity \eqref{eq:cubic}: (a) $w_+=1$; (b) $w_+=6$. Here $a=-1$, $b=10$, and the dashed vertical lines mark corresponding superkink velocity values.}
\label{fig:floquet_cubicSW}
\end{figure}

We also examined stability of the obtained solitary waves solutions using Floquet analysis and direct numerical simulations. In this case, the Floquet analysis also shows eventual emergence of spurious oscillatory instabilities that are similar to the ones we saw in the case of superkinks. As shown in Fig.~\ref{fig:floquet_cubicSW}, these instabilities are very mild: the maximum modulus, which increases as the superkink limit is approached, is bounded by $1.0006$ and $1.001$ in the two cases shown.

\begin{figure}
\centering
\includegraphics[width=\textwidth]{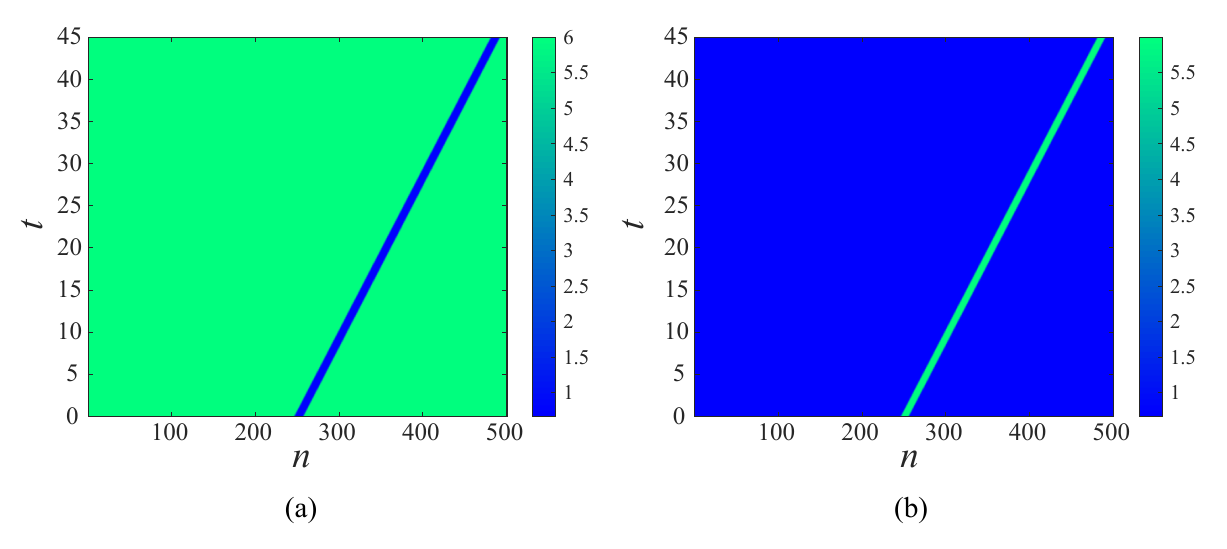}
\caption{\footnotesize Space-time strain evolution initiated by computed solitary wave solutions slightly below and slightly above the superkink limit, with velocities (a) $V=5.21749194749$ (a compressive wave); (b) $V=5.2174919476$ (a tensile wave). Here $a=-1$, $b=10$, $w_+=6$, $V_{SK}=7\sqrt{5}/3=5.21749194749951$.}
\label{fig:stabSW}
\end{figure}

Direct numerical simulations initiated by computed solitary waves show their robust propagation with velocity within $O(10^{-8})$ or less from the prescribed value and suggest that the waves are effectively stable, or at least long-lived, in the entire velocity range;
see Fig.~\ref{fig:stabSW} for representative examples.

\begin{figure}
\centering
\includegraphics[width=\textwidth]{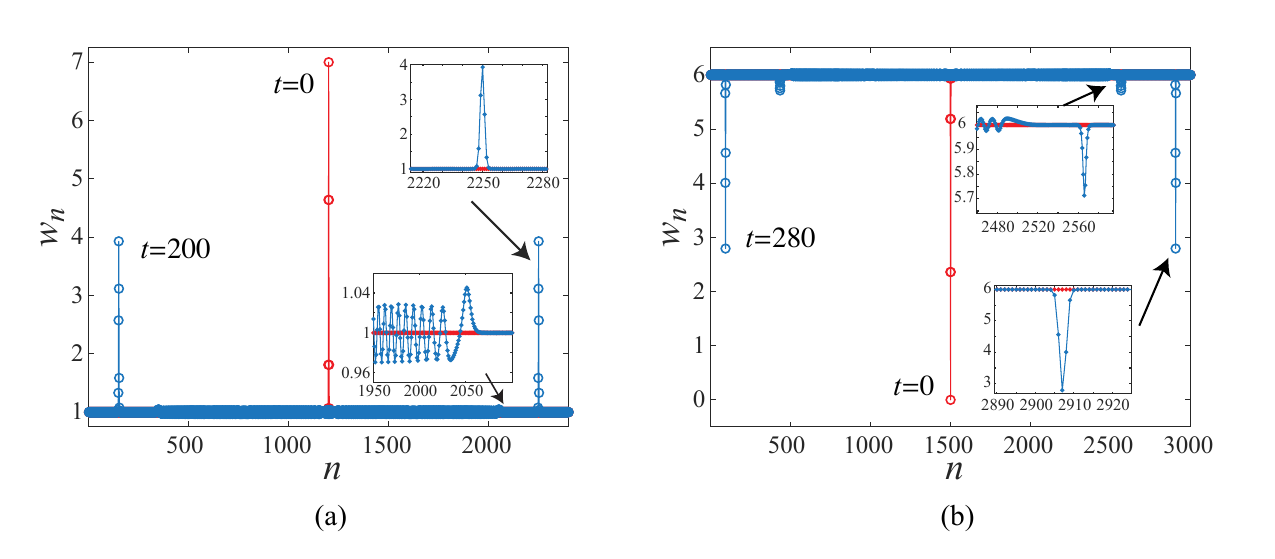}
\caption{\footnotesize Initial and final strain profiles in simulations with cubic nonlinearity \eqref{eq:cubic} and initial data \eqref{eq:GaussIC} with (a) $w_B=1$, $A=6$, $L=2400$; (b) $w_B=6$, $A=-6$, $L=3000$. Insets zoom in on the solitary and dispersive waves. Here $a=-1$, $b=10$.}
\label{fig:SWdyn}
\end{figure}

We also considered generic Gaussian-type initial conditions of the form
\beq
w_n(0)=w_B+A\exp[-(1/2)(n-L/2)^2], \quad \dot{w}_n(0)=0, \quad n=1, \dots, L
\label{eq:GaussIC}
\eeq
Using this initial data with various background strain $w_B$ and signed amplitude $A$ in simulations with free boundary conditions, we observed formation and steady propagation of both tensile (for $A>0$) and compressive (for $A<0$) solitary waves. Two examples are shown in Fig.~\ref{fig:SWdyn}.
In the first example, shown in Fig.~\ref{fig:SWdyn}(a), we set $A=6$ and $w_B=1$. One can see formation of two tensile solitary waves of the same form propagating in opposite directions with velocities $V=\pm 5.2494$. The waves are trailed by small-amplitude dispersive waves. In the second case (Fig.~\ref{fig:SWdyn}(b)), where we set $A=-6$ and $w_B=6$, there are two pairs of compressive solitary waves, the smaller-amplitude waves moving with velocities $V=\pm 3.811$ and the large-amplitude ones propagating with $V=\pm 5.025$, with dispersive waves trailing the smaller-amplitude solitary waves.

\section{Conclusions}
\label{sec:conclusions}
Classical studies of FPU-type systems involve weak nonlinearity, where the Hookean force-elongation relation of the linear theory is replaced by the simplest quadratic relation describing either hardening or softening of the mechanical response of the springs.
The main nonlinear effect in such setting is the emergence of solitary wave solutions parameterized by their velocities and stretching continuously over a semi-infinite range of velocities from the weak, strongly continuum near-sonic waves to the strongly discrete ones moving with arbitrarily large supersonic velocity. In this setting, a hardening nonlinearity produces tensile solitary waves while a softening nonlinearity generates their compressive analogs.

In this paper we considered the synthetic model containing hardening-softening nonlinearity, which can be represented by the simplest cubic force-strain relation. In other words, we considered a version of the classical Hamiltonian FPU problem with peculiar springs where a hardening response is taken over by a softening regime above a critical strain value. The resulting dynamic picture is expectedly more complex, with both tensile and compressive solitary waves simultaneously present, even if in different parameter ranges.

The proposed version of the FPU model was also shown to demonstrate a fundamentally new feature emerging as a result of the interplay between hardening and softening.  Thus, in addition to conventional solitary waves, such discrete system also supports non-topological and dissipation-free kinks. More precisely, we showed that in the proposed model, instead of growing without bound, the amplitude of both compressive and tensile solitary waves saturates around a velocity value which can be interpreted as a critical regime. Around this value of the parameter the compressive and the tensile solitary waves each converge to a configuration that can be seen as a bundle (or tandem) of infinitely separated kinks and antikinks. The infinite width of such a bundle suggests that the effective correlation length diverges and the increasingly flattening top or bottom of the near-critical solitary waves points towards the formation of a ``second  phase'', which can now coexist with the original ground state, or the ``first phase''.

The emerging picture is rather remarkable given that the elastic energy density remains convex within the range of strains involved in these solutions. The emergence of the ``second phase'' can be thus interpreted as a purely dynamical phenomenon, requiring a delicate interplay between kinetic and potential energy which are then conspiring to produce an effectively dynamic double-well structure. In this perspective solitary waves can be viewed as crossover features connecting sonic waves in both phases with the critical waves represented by stable supersonic kinks and antikinks. While the latter are fully nonlocal,  as they are conditioned by the limits at plus and minus infinity, they are non-topological,  in contrast to conventional sine-Gordon-type kinks,  as the limiting states are not separated by an elastic energy barrier. A striking feature of the proposed model is that both nonlocal kinks and local solitary waves   can move in a discrete setting with the same speed and without radiating lattice waves.

An interesting property of our hardening-softening version of the FPU model is that all the crucial features of the traveling wave solutions can be already captured by the simplest QC approximation, which, however, is not of a conventional KdV type and instead involves temporal dispersion.  The analytical transparency of the proposed QC model allowed us to corroborate and to rationalize theoretically various effects observed in our numerical investigation of the discrete model. Near superkink limit the observed agreement between the QC and discrete models was not only qualitative but also quantitative, which is not surprising in view of the critical nature of such regimes.

Finally, we mention that various localized traveling waves studied in this paper can be viewed as elementary bites of mechanical information that can be generated, delivered, and erased in periodic lattice metamaterials. Due to the presence of stress-sensitive repeating structural units, such metamaterials can be designed to exhibit complex mechanical response, in particular, to ensure that the mechanically triggered switching and actuation takes place at a predefined place and at a given level of stress.

\begin{section}*{Acknowledgements}
 The work of AV was supported by the NSF grant DMS-2204880. LT  acknowledges the support of the French Agence Nationale de la Recherche under the grant ANR-17-CE08-0047-02.
\end{section}

\end{document}